\documentclass{article}
\usepackage{geometry}
\usepackage[english]{babel}
\usepackage[font=small,labelfont=bf]{caption}
\usepackage{authblk}
\usepackage{float}
\usepackage[utf8]{inputenc}
\usepackage{listings}
\usepackage[ruled]{algorithm2e} 
\usepackage{amsfonts,amssymb,amsmath,amsthm}
\usepackage[shortlabels]{enumitem}
\usepackage{units}
\usepackage{tikz}
\usetikzlibrary{arrows,calc,positioning}
\usepackage{graphicx} 
\usepackage [autostyle, english = american]{csquotes}
\usepackage{bm}
\usepackage{longtable,multirow}
\usepackage[dvipsnames]{xcolor}
\usepackage [english]{babel}
\usepackage{comment}
\usepackage[T1]{fontenc}
\usepackage{pgfplots}
\pgfplotsset{compat=1.17}

\RequirePackage{natbib}
\setcitestyle{authoryear} 
\let\cite\citep

\usepackage{hyperref}

\MakeOuterQuote{"}
\title{Beyond the mean: Sequence analysis methods for clustering ordinal EMA data}

\author[1]{Tianyi Wang}
\author[1]{Anna L. Smith}
\author[2]{Jillian R. Silva-Jones}
\author[3]{Wendy Berry Mendes}
\author[4]{Lauren N. Whitehurst}

\affil[1]{Dr. Bing Zhang Department of Statistics, University of Kentucky}
\affil[2]{Department of Psychology, College of Arts and Sciences, University of Kentucky}
\affil[3]{Department of Psychology, Yale University}
\affil[4]{Departments of Psychology and Black Studies, Yale University}

\date{April, 2026}

\begin{document}

\maketitle

\begin{abstract}
Ecological momentary assessment (EMA) ratings are widely used in studies of behavioral and psychological phenomena to capture real-time data in subjects' real-world environments. 
Because the data are collected repeatedly over the study period, they provide rich longitudinal rating profiles for each individual. However, the number of observations per subject is often large, while both sample size and sampling intensity can vary substantially across individuals, which complicates the analysis. In some settings, simplified summaries of individual profiles, such as averages computed across the study period, are used for downstream analyses, including regression-style modeling. Although such summaries can be convenient, they may fail to fully capture dynamic temporal patterns present in the complete longitudinal profiles. 
To address this, we borrow measures from sequence analysis that capture individual-level patterns over time and then applied principal component analysis (PCA) followed by $K$-means clustering to identify unobserved latent groups of individuals with similar profiles. 
We test our approach using simulated data from a categorical functional regression model and compare its performance with two commonly used methods for detecting unobserved group structures: latent class analysis (LCA), and  latent transition analysis (LTA). Using EMA stress observations from a large sample of U.S. adults \citep{MYBPLAB2023, MYBPLab2025}, we identify distinct latent stress profile groups and show that they improve characterization of the impact on cognitive performance.
\end{abstract}


\section{Introduction}
Research studies across various fields, such as psychology, healthcare, and the social sciences, are leveraging ecological momentary assessment (EMA) methods, which data are collected under real-world environments to record study participants' behaviors, feelings, and experiences in real-time, within the natural context of their daily lives \citep[e.g.,][]{EMA_paper, paterson_armitage_turner_2023, browning_pinchak_calder_etal_2024}. 
This approach helps minimize recall bias and allows for the study of small-time-scale processes that may influence behavior in real-world environments \citep{shiffman_stone_hufford_2008}.
EMA data are often collected via electronic devices, such as mobile phone apps, at regularly scheduled intervals, and can include physiological sensors (e.g., heart rate, blood pressure) as well as self-report survey questions. 
As in traditional surveys, many measurements of interest are collected as categorical or ordinal (e.g., Likert-style) variables.
For example, (nominal) categorical response data can capture demographic information (e.g., race, gender, education), emotional states, or insurance status, while ordinal response scales can measure pain levels (e.g., 1 to 10), customer satisfaction, or perceived daily stress (e.g., 1 to 5), as we will discuss in detail below.

As a result of EMA data collection, researchers now have access to fine-scale longitudinal ordinal data, which can provide detailed insights into how these measures evolve and influence study outcomes over time.
Beyond characterizing general longitudinal trends, researchers are often interested in clustering individuals who exhibit similar patterns over time. 
For the motivating study we discuss below, there are competing theories in the literature regarding how individuals experience stress over time.
This evidence of natural clustering in individual stress profiles motivates us to consider whether EMA-collected daily stress ratings can be clustered into groups of individuals who exhibit similar patterns over time and whether these clusters are associated with a key study outcome, cognitive performance on a test of declarative memory.

In the statistics literature, methods for analyzing fine-scale variation over time, called functional data analysis \cite{wang_chiou_muller_2016}, have focused on continuous outcomes, although methods to accommodate categorical responses are being actively developed \citep{cfda}. \citet{EMA_paper} summarized a variety of modeling approaches that have been used to analyze EMA data. Depending on the purpose of the study and the structure of the data, it is not always necessary to include time as a covariate in the model, even when EMA data are collected at specific time points during the study period. For between-subjects analyses collected at predefined time points and analyzed using statistical methods such as Pearson correlation across individuals, the variables are typically assumed to be continuous outcomes, hence these methods are not suitable for ordinal or categorical variables. In addition, some studies focus on comparing event-related effects rather than between-subject differences. Analyses in these contexts often rely on event-based methods, typically within case-control designs. Analytical approaches for such designs include generalized estimating equations (GEE) \cite{GEE_1988}, dependent t-tests, and survival analysis, which are not designed for studies that do not include a control group. To investigate relationships involving both main effects and interactions among covariates, point process models can be employed. For example, thinned inhomogeneous Poisson processes are well-suited for sparse event data and allow modeling of event intensity over time. While useful for estimating event rates, they are unsuitable for repeated measures data and cannot capture latent population structures. More flexible approaches, such as population-averaged (PA) and subject-specific (SS) models, offer improvements: PA models estimate marginal effects across the population, while SS models account for within-subject variation using random effects. However, both assume a homogeneous population and cannot uncover latent subgroups.

As an alternative, we borrow tools from sequence analysis which are explicitly designed to study longitudinal categorical data and were originally proposed by biologists to study patterns in DNA sequences \citep{SA_New_method_Old_idea_1995}.
We group individuals with similar stress patterns via principal components analysis (PCA) of our sequence analysis measures, and  apply $K$-means clustering algorithms \cite{PCA_and_K_means, Algroithm_Cluster_1988, Fuzzy_C_means_1973}.
As a result, our approach uses tools that are familiar and accessible to many data scientists and applied researchers, while effectively leveraging relevant information from high-frequency EMA data.

To validate our approach, we demonstrate that it can recover cluster memberships accurately and is more stable than alternative na\"{i}ve clustering approaches, including latent class analysis (LCA) and latent transition analysis (LTA), across simulated data from a sophisticated categorical functional data model.
For our motivating study, we examine whether our estimated clusters describe distinct stress patterns and compare how well stress cluster membership predicts cognitive performance in a binomial mixed-effects model, relative to reasonable alternative predictor variables: (1) overall mean stress (averaging over time), (2) a selection of (correlated) stress summary statistics (e.g., the mean, standard deviation, mode, measures from the sequence analysis and categorical time series literature), and (3) PCA loadings for the stress summary statistics. We also discuss how to choose the number of principal components, the number of clusters, and examine an alternative clustering algorithm that allows for overlap among the estimated clusters.

The paper is structured as follows: Section ~\ref{sec:motiv_study} introduces the motivating study, provides theoretical background on stress and its potential health links, and the MyBPLab 2.0 dataset \citet{MYBPLAB2023, MYBPLab2025}. Section ~\ref{sec: existing_app} discusses the mixed-effects modeling framework within the MyBPLab setting and introduces two common approaches for uncovering unobserved latent groups, namely LCA and LTA. Section ~\ref{sec:method} describes relevant tools from sequence analysis and presents our PCA-based approach. Section ~\ref{sec:sim_cfreg} applies the PCA-based method to simulated data under four settings and compares its performance to LCA and LTA. Section ~\ref{sec:realdata_app} presents the results of five mixed-effects models applied to the MyBPLab data and compares their performance. Finally, Section ~\ref{sec:disc} discusses limitations of our approach and outlines directions for future research.

\section{Motivating study: EMA-collected stress ratings and cognitive performance}\label{sec:motiv_study} 
Stress is a common experience in our daily lives but undue stress and poor stress responses have been linked to increased risk of experiencing or exacerbating many adverse health outcomes, such as heart disease, cancer, asthma, and declining cognitive performance, including memory \cite{Stress_effects_health_2008,ChronicStress2017}. 
 During a stressful event, cognitive processes prioritize evaluating information directly relevant to the stressful event, reducing cognitive resources available for episodic memory  \cite[the recall of explicit facts;][]{Schwabe_recall_explicit_fact_2022}, for retrieving previously acquired memories \cite{Emotional_memories_buchanan_2006, stress_memory_affect_smeets_2008}, for goal-directed learning \cite{goal_directed_stress_Fournier_2017,Goal_directed_Quaedflieg_2019}, for memory updating \cite{Memory_updating_Dongaonkar_2013, Mem_updating_Nitschke_2019}, and for the generalization of memories \cite{Generalization_mem_Dandolo_and_Schwabe_2016}.
This body of research focuses on how memory is impacted by co-occurring (or recent) stressful events at a static point in time, often called acute stressors.
However, experiencing an acute stressor is a rare phenomenon and recent theoretical advances \citep[e.g.,][]{brosschot_verkuil_thayer_2018} highlight the potential impact of persistent or pervasive stress, which is not bounded by time and often leads to the most significant health problems. EMA measures provide small-time-scale measures of perceived daily stress which allow us to examine how \textit{patterns} of stress over time may impact memory.
Our methodological approach seeks to uncover whether any common patterns exist across individuals, i.e., latent subgroups. To motivate this approach, we briefly summarize competing theories of how stress responses vary over time in the following section.

\subsection{Competing theories of stress activation}
In stress science, researchers have proposed competing theories for how individuals react to stressors and how their stress responses impact broader health  \cite{epelMoreFeelingUnified2018}. 
Broadly, these theories describe how systems of the body -- such as the autonomic, central nervous, neuroendocrine and immune systems -- respond to challenges in order to maintain homeostasis in the body.
When an individual experiences a stressor, these adaptive systems are regulated upward or downward, to bring the body back to homeostasis, allowing the individual to cope with challenges that threaten its survival.
This adaptation process has been called ``allostasis'' or ``stability through change'' \cite{sterling1988handbook}.

The allostatic load theory of stress argues that if these adaptive systems are overstimulated, activated too frequently, or fail to downregulate after the stressful event, then the cumulative toll of stress on the physiological system results in “allostatic load”, which can lead to the harmful disease outcomes mentioned above \cite{mcewen_1998}.
The cognitive activation theory of stress (CATS), builds on this idea and formulates the stress response as a general alarm within the body’s homeostatic system, which increases the body’s level of allostatic activation from one level of activation to more activation \cite{ursin_eriksen_2004}.

On the other hand, the generalized unsafety theory of stress (GUTS), pulling from neurobiological and evolution-theoretical research, argues that for some groups, particularly those that are vulnerable or not feeling safe, ``the stress response is a default response that is always `on’ but inhibited by the prefrontal cortex when safety is perceived'' \cite{brosschot_verkuil_thayer_2018}. 
From this perspective, the prolonged stress responses that have been linked to adverse health outcomes are the result of nonspecific and largely unconscious perceived unsafety rather than the faulty firing of allostatic systems in response to specific stress events.

Translating these theories to individual-level observations of stress, CATS implies that daily self-reported stress ratings are most frequently low, with potentially stressful situations showing up less frequently, as days or instances with high ratings. GUTS implies precisely the opposite:  frequently higher (moderate) stress ratings, with low stress ratings being less frequent. 
This is consistent with the deep rest model, which leverages GUTS to argue that the baseline state is moderate stress arousal.  Under this model, the biological embodiment of stress depends on safety signaling, which is tied to social, cultural, and political factors and interventions like contemplative practice \citep{crosswellDeepRestIntegrative2024}. These theories demonstrate that scientists' understanding of how stress is patterned over time is of key interest.
It motivates us to quantify to what extent stress patterns over time are clustered across individuals, 
and how these shared stress profiles impact broader health outcomes. In Section ~\ref{sec:realdata_app}, we estimate shared stress profiles from EMA data and examine its impact on cognitive performance.
Our goal is not to evaluate support for either theory of stress, but rather to demonstrate that characterizing how stress profiles are patterned over time can improve our understanding of its impacts on overall health.

\subsection{Data Structure}\label{sec:data_cleaning}
We examine data collected by \citet{MYBPLAB2023, MYBPLab2025} in a large, geographically diverse study of stress and its association with blood pressure (BP) reactivity. Participants were able to join the study with an eligible smartphone or watch (i.e., a Samsung phone with a built-in optic sensor) after downloading the MyBPLab app through the U.S. Google Play Store. Study data included physiological measurements, such as BP and heart rate (HR), along with self-reported emotional, psychosocial, and behavioral data. Stress levels were measured by the following prompt : "Have you experienced any particularly stressful events since your last check-in?" A response of `Yes' is interpreted as an acute stress experience, while response of `No' would be interpreted as the absence of an acute stress exposure. Our focus will be on daily stress patterns over time, retaining only records with a `No' response, which corresponds to 84\% of stress responses. A follow-up question collected participants' self-reported  levels of stress using a 5-point Likert-scale.  Participants were asked to record how much they agree with the statement, "I feel stressed, anxious, overwhelmed," using a five-point Likert scale: not at all (1), a little bit (2), somewhat (3), moderately (4), and extremely (5). We let $X_{it}$ represent the stress score for individual $i$ at time $t$. For simplicity, we index time discretely, with $t=1, 2, … T_i$ representing each of individual $i$’s subsequent interactions with the data collection app.

\begin{figure}[h]
      \centering                        
      \includegraphics[width=0.8\textwidth]{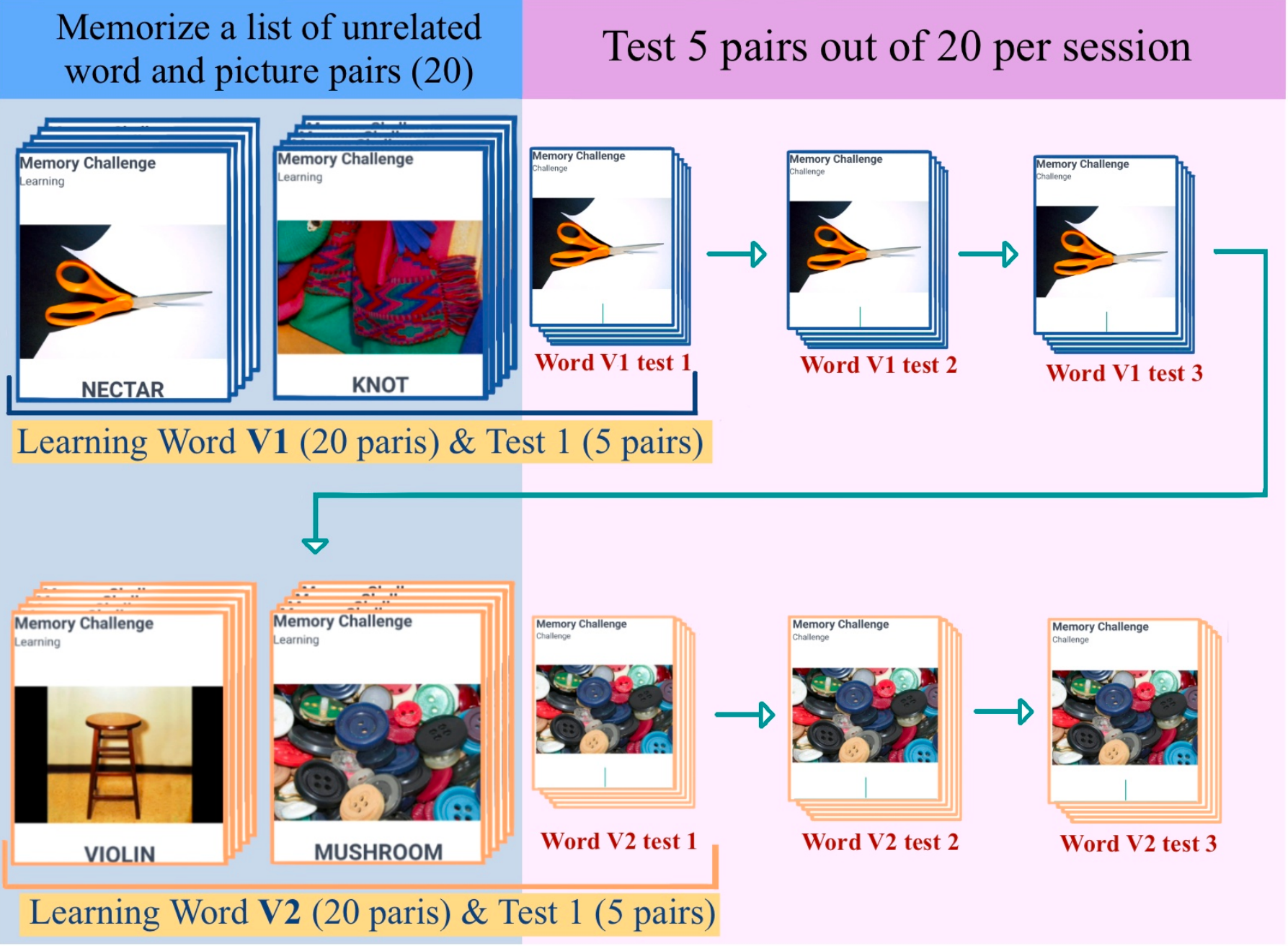}
      \caption[]{\small Memory test from MyBPLab data \citep{Sliva_UK_2025_EMA, Gilmore_2024_EMA_memory}. There are two photo-word sets; each learning session is followed by three test sessions on different days.}
      \label{fig:memory_test}
\end{figure}
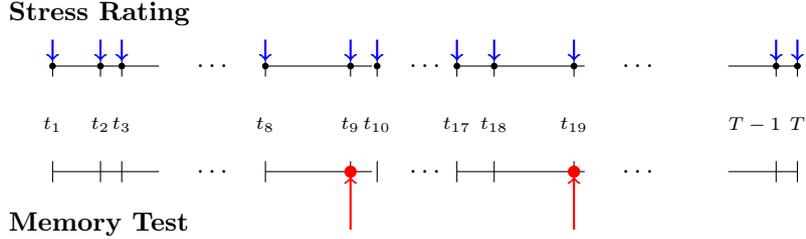
\begin{figure}[htbp]
  \centering
    \begin{tikzpicture}[scale=0.7]
        \def\endPoint{14} 
    
        \draw (0,0) -- (2,0); 
        \node at (3,0) {$\ldots$}; 
        \draw (4,0) -- (6,0); 
        \node at (7,0) {$\ldots$}; 
        \draw (7.6,0) -- (10,0); 
        \node at (11,0) {$\ldots$}; 
        \draw (12.7,0) -- (\endPoint,0); 

        \draw (0,-2) -- (2,-2); 
        \node at (3,-2) {$\ldots$}; 
        \draw (4,-2) -- (6,-2); 
        \node at (7,-2) {$\ldots$}; 
        \draw (7.6,-2) -- (10,-2); 
        \node at (11,-2) {$\ldots$}; 
        \draw (12.7,-2) -- (\endPoint,-2); 
    
         \foreach \x in {0, 0.9, 1.3, 4, 5.6, 6.1, 7.6,8.3, 9.8, 13.6, 14} {
             
            \draw (\x,0.2) -- (\x,-0.2);
            \draw (\x,-1.8) -- (\x,-2.2); 
     
            \filldraw[black] (\x,0) circle (1.5pt);
            \draw[->, thick, blue] (\x,0.5) -- (\x,0.1);
            \ifdim\x pt=5.6pt
                \draw[->, thick, red]  (\x,-3.1) -- (\x,-2.1);
                \filldraw[red] (\x,-2) circle (3pt); 
            \fi
            
            \ifdim\x pt=9.8pt
                \draw[->, thick, red] (\x,-3.1) -- (\x,-2.1); 
                \filldraw[red] (\x,-2) circle (3pt); 
            \fi
            
         }
         \node[anchor=north, font=\scriptsize] at (0,-0.8) {$t_{1}$};
         \node[anchor=north, font=\scriptsize] at (0.9,-0.8) {$t_{2}$};
         \node[anchor=north, font=\scriptsize] at (1.3,-0.8) {$t_{3}$};
         \node[anchor=north, font=\scriptsize] at (4,-0.8) {$t_{8}$};
         \node[anchor=north, font=\scriptsize] at (5.6,-0.8) {$t_{9}$};
         \node[anchor=north, font=\scriptsize] at (6.1,-0.8) {$t_{10}$};
         \node[anchor=north, font=\scriptsize] at (7.6,-0.8) {$t_{17}$};
         \node[anchor=north, font=\scriptsize] at (8.3,-0.8) {$t_{18}$};
         \node[anchor=north, font=\scriptsize] at (9.8,-0.8) {$t_{19}$};
         \node[anchor=north, font=\scriptsize] at (13.2,-0.8) {$T-1$};
         \node[anchor=north, font=\scriptsize] at (14,-0.8) {$T$};
    
        \node[anchor=west] at (-1,1) {\textbf{Stress Rating}};
        \node[anchor=west] at (-1,-3) {\textbf{Memory Test}};
    
    \end{tikzpicture}
  \caption{\small An example timeline for memory tests and stress rating observations. Red and blue arrows indicate timepoints at which memory tests and stress ratings are recorded, respectively.}
  
  \label{fig:mem_stress_time} 
  
\end{figure}

While participants may record stress ratings multiple times per day, the memory test is accessible less frequently and is scheduled to occur only once within predefined time windows, to ensure sufficient encoding time for the memory task. Additionally, the number of stress ratings between subsequent memory tests varies across individuals and pairs of tests (Figure ~\ref{fig:mem_stress_time}) . Among individuals with more than 25 stress rating observations and at least one memory test record ($n = 1215$), the average and median number of memory tests is 2.82 and 2, respectively, while the average and median number of stress ratings is 71.75 and 47, respectively.

\section{Existing approaches}\label{sec: existing_app}
Our goal is to explain how participants' self-reported stress levels impact their cognitive performance on the memory tests, while incorporating the high frequency of stress reports available in the data, and to evaluate whether the data support clusters, or common patterns, of stress ratings over time across participants, as indicated by the competing theories of stress activation.
One simple approach is to specify a cross-sectional model that summarizes each individual's observations across time. The mean is the most commonly used summary statistic, although other measures, such as variability or the most recent preceding observation, can also be used \cite{Repeated_measure_mean_2018,Variability_Longitudinal_2012_Ellliott}. The selected summary statistic is then used as a covariate in a regression model for the outcome variable of interest. Use of these summary statistics results in a substantial loss of information about potential stress profile groups and individuals’ stress patterns over time.

The memory test consisted of two sections, learning and testing, in which participants studied two versions of 20 unrelated word–photo pairs during the learning phase, and in each subsequent test section were presented with five photos at a time and asked to type the correct word paired with each photo, with the first test conducted on the same day as the learning session and the second and third tests administered on separate subsequent days to assess retention over time, see Figure ~\ref{fig:memory_test}. We define $M_{ijt}$ as a binary variable which records participant $i$'s response for the $j$th word-photo pair at time $t$. Our analysis will focus on $Y_{it} = \sum_{j=1}^5 M_{ijt}$, which represents the total number of correctly identified word-photo pairs(out of five) for individual $i$ at time $t$. We treat $M_{ijt}$ as independently distributed according to a Bernoulli distribution with parameter $p_{it}$, which implies that the outcome variable, $Y_{it}$ is an independent draw from a Binomial distribution. In our motivating study, this corresponds to assuming independence across word-photo pairs within and across memory tests completed by the same participant. Our Binomial regression model posits that $p_{it}$, the probability that individual $i$ will correctly respond to a memory challenge at time $t$, is linearly related to the individual's average stress score, where we average across all time points, and also controls for common demographic variables \citep{mccullagh_2019}.
The generalized linear mixed model (GLMM)  is illustrated below:

\begin{align}  
    M_{ijt} &\overset{\perp}{\sim} \text{Bernoulli}(p_{it}) \nonumber\\
    Y_{it} = \sum_{j=1}^5 M_{ijt}  &\overset{\perp}{\sim} \text{Binomial}(5,p_{it}) \label{eq:glmer_model}\\
    \text{logit}(p_{it}) &= \beta_{0m} +\beta_{1m}\bar{X}_{i \cdot}+\sum_l \gamma_l D_{il}+ u_{it} \nonumber
\end{align}
where $\bar{X}_{i \cdot} = \frac{1}{T_i} \sum_{t=1}^{T_i} X_{it}$ represents the mean stress level across time for individual $i$. $D_{il}$ are demographic variables, including age, gender, education, tobacco use, BMI and race. $\beta_{0,m}, \beta_{1,m}$ and $\gamma_l$ are regression coefficients \cite{Book_2016_GLMM_R_faraway}. In \texttt{R}, we can use \texttt{weights = $n\times 5$} to assign weights to the response variable since each memory task consists of five test questions. The $u_{it}$ are random effects to account for dependence across memory tasks completed by the same individual, where $ u_{it} \sim \text{Normal}(0,\sigma_i)$. \  

Alternatively, since we are interested in uncovering common patterns of stress over time, we would like to move beyond $\bar{X}_{i \cdot}$, which collapses all stress measurements over time, and instead create a proxy covariate from a latent clustering method, such as LCA and LTA. LCA identifies clusters of similar individuals or cases from a set of related observed binary variables or items \cite{hagenaars_applied_2002}. For example, standard screener surveys for depression include multiple items (potential symptoms) for which patients record the absence or presence of each depressive symptom.  LCA of these data identify qualitatively distinct groups of individuals who experience similar sets of depressive symptoms, implying clinical subtypes of depression \citep{HeterogeneityPostpartumDepression,mournet_kleiman_2024}.
 Longitudinal extensions of LCA do exist, but typically place severe limits on the number of time points considered for each individual, often to less than three to preserve identifiability of the parameters  \cite{hagenaars_applied_2002}. In our study, individuals record their stress levels at roughly 72 time points, on average.  However, we could naively consider the stress measurement at each time $t$ as a separate covariate. The LCA model then requires assuming conditional independence of the stress scores at each time point and is specified as follows: 
\begin{align*}
    f(X_{i1}, \dots,X_{iT}) = \sum^K_{k = 1}p_k \prod^T_{t=1}\prod^{5}_{c =1}\pi_{tkc}^{I(X_{it} = c)}
\end{align*}
where $p_k = P(Z_i = k)$ are `prior' probabilities of latent class memberships,
$\pi_{tkc} = P(X_{it}=c|Z_i = k)$ denotes the class-conditional probability that individual $i$'s stress rating at time $t$ is observed as $c$, given that individual $i$ belongs to latent class $k$ and $Z_i \in \{1, \dots K\}$ represents individual $i$'s latent class membership for $c = 1, ...5$. The model requires that $\sum^K_{k=1}p_{k} = 1$;  since the stress scores are recorded as categorical variables, we have the additional constraint that $\sum^{5}_{c=1}\pi_{tkc} = 1$, for each time point, $t$, and latent class, $k = 1, \dots, K$ \cite{linzerPoLCAPackagePolytomous2011}. LTA on the other hand, is designed for longitudinal data but allows individuals to transition between latent classes at each time point. This method also been widely used in psychological and mental health sciences to study states changing over time\cite{LTA_psy_study_2010,LTA_ALSPAC_2017,LTA_bully_data}. While there is no universal threshold for the number of time points that latent transition analysis (LTA) can accommodate, studies typically prefer a small number of time points with substantial real-time separation; for example, \citet{LTA_bully_data} measured outcomes across three semesters, and \citet{maldonado-molinaPatternsSubstanceUse2007} examined the progression of substance use among Hispanic youth over two years (1998 and 1999). For our motivating study, competing theories of stress activation suggest that there may be different patterns of stress rating profiles across individuals, but the literature does not suggest that individuals switch between activation patterns over time as LTA assumes. The LTA model can be written as: 
\begin{align*}
    f(X_{i1}, \dots, X_{iT}) = \sum_{\bm{Z_i}} \gamma_{Z_{i1}} \prod^T_{t=2} \gamma_{Z_{it} | Z_{i,t-1}} \prod^T_{t=1} \phi_{X_{it}|Z_{it}}
\end{align*}

where $Z_{it}$ is the latent state membership for individual $i$ at time $t$, $\phi_{X_{it}|Z_{it}} = P(X_{it}|Z_{it})$ is the class-conditional probability of the $i$th individual's response at time $t$, $\pmb{Z_i} = (Z_{i1},\dots, Z_{iT})$ is a latent process that is assumed to follow a first-order Markov chain with state space $\{1,\dots, K\}$, $\gamma_{Z_{1t}}$ is the initial probability of the latent process, and $\gamma_{Z_{it}|Z_{it-1}} = P(Z_{it}|Z_{it-1})$ is the transition probability \cite{Latent_Markov_LMEST_R_LTA}. Both LTA and LCA are  computationally computationally intensive (and often infeasible) as the number of time points increases and cannot handle varying numbers of time points across individuals. In practice, real-world data are often collected over long periods with high-frequency assessments. Thus, we turn to more flexible approaches that perform well within familiar regression frameworks.

\section{Building better predictors: Sequence analysis, PCA, and $K$-Means}\label{sec:method}
Sequence analysis 
explicitly captures categorical patterns over time, and is popular in social science applications \cite{SA_New_method_Old_idea_1995}. In particular, a sequence index plot is an effective tool for analyzing sequences derived from coarsely measured events, such as employment status or educational attainment. This plot summarizes sequencing information, including the duration and frequency of each state across individuals. In such contexts, the sequences typically have fewer time points, lower complexity, and are measured over fixed intervals. Additionally, certain transitions, such as backward movement in educational level, are rare or even structurally restricted. A notable limitation is that only a relatively small number of individuals (typically around 200) can be displayed in a single graph to avoid excessive overplotting and ensure interpretability \cite{SA_past_present_future_2022}. 
\begin{figure}[H]
      \centering                        
      \includegraphics[width=1\textwidth]{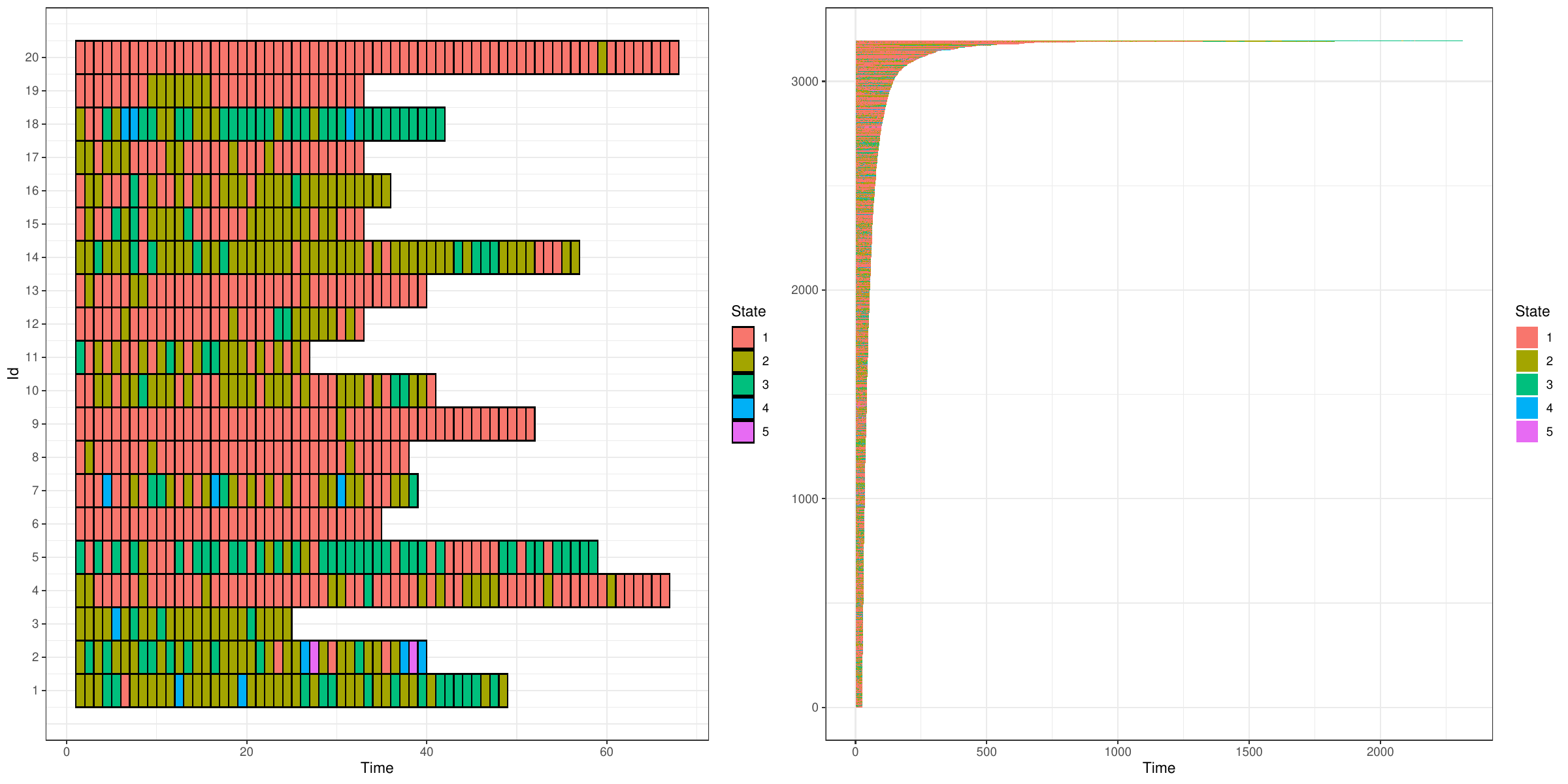}
      \caption[]{\small Sequence index plot (left: 20 individuals, $\max_i T_i \approx 70 $; right: all individuals, $\max_i T_i \approx 1750 $). Each color represents a distinct stress state. This figure is created using the \texttt{plotData} function from the \texttt{cfda} package \cite{cfda}. }
      \label{fig:seqindex}
\end{figure}
 In our data, each horizontal row represents an individual's stress states over time. Each block indicates a specific stress level at a given time point, with different colors corresponding to different levels of stress (1 to 5). The entire row displays an individual's stress level changes across time. The plot on the left displays data from 20 randomly selected subjects, while the plot on the right shows longitudinal stress data for all individuals. As presented in Figure ~\ref{fig:seqindex}, the sequences show substantial variability across individuals, with differences in the number of time points and the frequency of transitions between stress states. As a result, the high variability and density of changes in our data can overwhelm traditional sequence index plots, making it difficult to extract meaningful patterns at the population level. Therefore, while sequence index plots offer intuitive visual summaries in conventional sequence analysis (SA) applications, they may not be well-suited for analyzing the complex, high-frequency stress dynamics observed in our study, but the core idea behind SA remains valuable. In particular, SA emphasizes capturing individual-level temporal patterns, which aligns with our goal of characterizing how trajectories evolve over time.
 
To leverage this insight, we draw upon related methodological work designed to handle more complex categorical trajectories. For example, \citet{cat_TS_2010_Elzinga} noted that the complexity between two sequence of responses can vary a lot depending on the ordering of the data. This insight inspired our observation that even if two individuals have the same mean stress level, their overall stress patterns can still differ markedly. Consider the three sequences below, all of which share the same mean. 
\[
\begin{array}{lll}
    A:123123123 & B: 111222333 & C: 133221132\\
\end{array}
\]

Compared to Sequences A and B, Sequence C exhibits a more random pattern, whereas both Sequences A and B display clear, repeated structures. This observation motivates us to introduce a lag variable, defined as $L_{it} = X_{it}-X_{i(t-1)} $, where $t \in \{2,\dots,T\}$, which represents the difference between consecutive time points. To incorporate this idea into our analysis, we draw on approaches based on the probability of specific responses by \citet{cat_TS_2010_Elzinga} and the characterization of ordered state paths by \citet{cfda}. Accordingly, we include the following summary measures, the mean lagged stress, denoted by $\bar{L}_{i\cdot}$, and the standard deviation of lag, $SD(L_{i})$, both of which capture the overall magnitude and variability of changes between time points. Finally, we compute the proportion of lag values equal to zero defined as $\frac{\sum^T_{t=2}I(L_{it}=0)}{T-1}$, which serves as a measure of temporal stability within each individual's stress sequence.

Building on these individual-level features, we adopt a strategy inspired by SA to group individuals with similar stress trajectories over time. TO incorporate our measure of sequence complexity and standard summaries of stress, we apply PCA to summarize and reduce the dimensionality of these individual-level features, followed by $K$-means clustering to identify groups of individuals with similar stress patterns \citep{PCA_2022_Green, Algroithm_Cluster_1988}.
%
%
\section{Simulation Study: Recovery of underlying latent profiles}\label{sec:sim_cfreg}
In this section, we introduce the simulation study by first describing the data-generating process and simulation settings. We then present the algorithms for applying our PCA-based approach and benchmark alternatives, LCA and LTA, in Section 5.2, followed by the results from the three approaches in Section 5.3.

\subsection{Generative model: Categorical Functional Regression}
In order to study how well our proposed approach can identify underlying patterns from longitudinal ordered data, we generate grouped data from a sufficiently complex model. \citet{cfda} defines a categorical functional data model which describes subject-specific trajectories as realizations of a stochastic jump process evolving in continuous time over a finite set of categorical states, and is implemented in the \texttt{cfda} package. We construct $C\times C$ transition matrices $L_k$ for underlying latent class $k$ where $ k = 1,\; \dots,\; K$. For each matrix $L_k$, the element $l_{c_p,c_q}=P(X_t = c_p\longrightarrow X_{t+1}=c_q)$ represents the probability of transitioning from state $c_p$ at time $t$ to state $c_q$ at time $t+1$. In this simulation study, we considered four settings, each comprising $K$ transition matrices $L_k$ for $ k = 1,\; \dots,\; K$.

Setting 1 is LCA-friendly, as class differences are primarily driven by state distributions rather than transition dynamics, and the weak temporal dependence aligns with the local independence assumption of LCA, as reflected by equal transition probabilities (0.25) across all states in the first class. The second class exhibits a preference for the first state (low or no stress), whereas the third class prefers states 2 and 3 (somewhat to moderate stress). Setting 3 and 4 are extensions of Setting 1. In Setting 3, noise is introduced to Class 2 and Class 3, while Class 1 remain unchanged. In Setting 4, greater overlap introduced to Class 2 and Class 3, again keeping Class 1 unchanged, thereby increasing both state overlap and transition variability. Setting 2 is LTA-friendly, characterized by a very strong transition pattern, with the matrix featuring a fixed deterministic cycle across states in the order $1 \rightarrow 2 \rightarrow 3 \rightarrow 4 \rightarrow 5 \rightarrow 1$. Classes 2 and 3 remain the same as in Setting 1. Notably, $L_1$ in Setting 2 does not follow a continuous-time Markov process, as it is based on a deterministic transition structure. The exact transition matrices for each setting are provided in Table ~\ref{tab:generator_MAX_set}.

\begin{table}[h]
\centering
\begin{tabular}{|c|c|c|}
 \hline
 \textbf{$L_1$} & \textbf{$L_2$}& \textbf{$L_3$} \\
 \hline
\multicolumn{3}{|c|}{\textbf{Setting 1: LCA-friendly}}\\
 \hline
$ 0.25(\bm{1} - I)$  &
 $\begin{array}{c}
  \vphantom{\begin{bmatrix}0\end{bmatrix}} \\
  \begin{bmatrix}
    0 & 0.25 & 0.25 & 0.25 & 0.25 \\
    0.7 & 0 & 0.1 & 0.1 & 0.1 \\
    0.7 & 0.1 & 0 & 0.1 & 0.1 \\
    0.7 & 0.1 & 0.1 & 0 & 0.1 \\
    0.7 & 0.1 & 0.1 & 0.1 & 0
  \end{bmatrix}\\
  \vphantom{\begin{bmatrix}0\end{bmatrix}}
 \end{array}$ & 
  $\begin{array}{c}
  \vphantom{\begin{bmatrix}0\end{bmatrix}} \\
  \begin{bmatrix}
    0 & 0.4 & 0.4 & 0.1 & 0.1 \\
    0.1 & 0 & 0.7 & 0.1 & 0.1 \\
    0.1 & 0.7 & 0 & 0.1 & 0.1 \\
    0.1 & 0.4 & 0.4 & 0 & 0.1 \\
    0.1 & 0.4 & 0.4 & 0.1 & 0
  \end{bmatrix}\\
  \vphantom{\begin{bmatrix}0\end{bmatrix}}
 \end{array}$ \\
\hline
\multicolumn{3}{|c|}{\textbf{Setting 2: LTA-friendly} }\\
\hline
 $\begin{array}{c}
  \vphantom{\begin{bmatrix}0\end{bmatrix}} \\
  \begin{bmatrix}
    0 & 1 & 0 & 0 & 0 \\
    0 & 0 & 1 & 0 & 0 \\
    0 & 0 & 0 & 1 & 0 \\
    0 & 0 & 0 & 0 & 1 \\
    1 & 0 & 0 & 0 & 0
  \end{bmatrix}\\
  \vphantom{\begin{bmatrix}0\end{bmatrix}}
\end{array}$ &
 $L_2$ from Setting 1 & 
$L_3$ from Setting 1
 \\
\hline
\multicolumn{3}{|c|}{\textbf{Setting 3: LCA-friendly + Noise}}\\
\hline

 $L_1$ from Setting 1&
 
 $\begin{array}{c}
  \vphantom{\begin{bmatrix}0\end{bmatrix}} \\
  \begin{bmatrix}
    0 & 0.25 & 0.25 & 0.25 & 0.25 \\
    0.7 & 0 & 0.02 & 0.16 & 0.12 \\
    0.7 & 0.08 & 0 & 0.12 & 0.1 \\
    0.7 & 0.21 & 0.05 & 0 & 0.04 \\
    0.7 & 0.09 & 0.18 & 0.03 & 0
  \end{bmatrix}\\
  \vphantom{\begin{bmatrix}0\end{bmatrix}}
 \end{array}$ & 
 $\begin{array}{c}
  \vphantom{\begin{bmatrix}0\end{bmatrix}} \\
  \begin{bmatrix}
    0 & 0.4 & 0.4 & 0.07 & 0.13\\
    0.2 & 0 & 0.7 & 0.05 & 0.05 \\
    0.08 & 0.7 & 0 & 0.12 & 0.1 \\
    0.19 & 0.4 & 0.4 & 0 & 0.01 \\
    0.05 & 0.4 & 0.4 & 0.15 & 0
  \end{bmatrix}\\
  \vphantom{\begin{bmatrix}0\end{bmatrix}}
 \end{array}$ \\
 
\hline
\multicolumn{3}{|c|}{\textbf{Setting 4: More Overlap} }\\
\hline
 $L_1$ from Setting 1 &
 
 $\begin{array}{c}
  \vphantom{\begin{bmatrix}0\end{bmatrix}} \\
  \begin{bmatrix}
    0 & 0.25 & 0.25 & 0.25 & 0.25 \\
    0.9 & 0 & 0.05 & 0.04 & 0.01 \\
    0.8 & 0.05 & 0 & 0.1 & 0.05 \\
    0.8 & 0.03 & 0.08 & 0 & 0.09 \\
    0.8 & 0.15 & 0.03 & 0.02 & 0
  \end{bmatrix}\\
  \vphantom{\begin{bmatrix}0\end{bmatrix}}
 \end{array}$ &

 $\begin{array}{c}
  \vphantom{\begin{bmatrix}0\end{bmatrix}} \\
  \begin{bmatrix}
    0 & 0.1 & 0.05 & 0.05 & 0.8 \\
    0.6 & 0 & 0.05 & 0.05 & 0.3 \\
    0.5 & 0.05 & 0 & 0.1 & 0.35 \\
    0.3 & 0.12& 0.08 & 0 & 0.5 \\
    0.8 & 0.15 & 0.03 & 0.02 & 0
  \end{bmatrix}\\
  \vphantom{\begin{bmatrix}0\end{bmatrix}}
 \end{array}$ \\
 \hline
\end{tabular}
\caption{\small Transition matrix settings: The matrices $L_1$ to $L_3$ represent the transition matrices for latent classes 1 through 3. Each is a $C \times C$ matrix, where the entry $l_{c_1, c_2}$ denotes the transition probability from level $c_1$ to level $c_2$ for $c_j \in \{1, \dots, C\}$. For example, in Setting 1 for matrix $L_2$, the entry $l_{2,1} = 0.7$ represents a 0.7 probability of transitioning from level 2 to level 1  }
\label{tab:generator_MAX_set}
\end{table}

\begin{figure}[h]
      \centering                        
      \includegraphics[width=.7\textwidth]{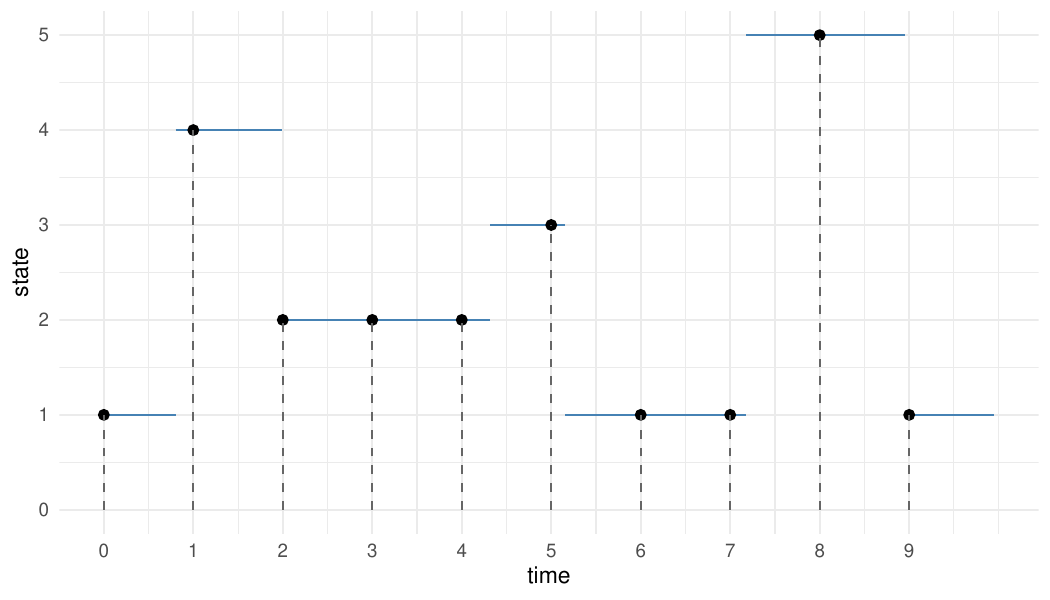}
      \caption[]{\small Discretization of continuous time: The blue horizontal lines represent the state values and their durations, while the solid black dots indicate the state at specific discrete time points marked by the dashed vertical lines. }
      \label{fig:locf}
\end{figure}

To mimic the MyBPLab data, we use $C = 5$ ordinal levels. For evaluation, we simulate stress trajectories over $T$ = 44 time points, which corresponds to the median number of check-ins in the MyBPLab study among individuals who had 25 or more non-missing stress observations, while varying the total number of individuals ($N = 600, 900, 1200, 1500$). The initial distribution of states was calculated using MyBPLab data \cite{MYBPLAB2023, MYBPLab2025}, selecting individuals with 44 or more valid records. Time spent in each state was uniformly set to 1, indicating the duration for each state until it moves to the next state is 1 time unit. By default, the functional data are simulated in continuous time, which we subsequently convert into discrete time points to better reflect the structure of the original dataset. A discrete time variable was created by taking the ceiling of the original continuous time. If the gap between the ceiling values for two continuous time points was greater than one unit of time (e.g, 2 and 5), then the missing discrete time points were added, and the corresponding state values of those time points were filled using the last observation carried forward algorithm. In this example, the state value at discrete time 2 would be used to fill times 3 and 4.  Figure ~\ref{fig:locf} plots an example subset of simulated states in continuous time for one individual, with vertical dashed lines indicating state entries corresponding to discrete time points. Table ~\ref{tab:generate_markov} provides the numeric values associated with Figure ~\ref{fig:locf}; notably, the state transitions occur according to the continuous time variable.

\begin{table}[h]
\centering
\begin{tabular}{ |c||c|c|c| c| }
 \hline
 ID & discrete time & continuous time & state & LOCF state\\
 \hline
 1 & 0 & 0.0000000 & 1 & 1\\
 1 & 1 & 0.8009673 & 4 & 4\\
 1 & 2 & 1.9858932 & 2 & 2\\
 1 & 3 & NA & NA & 2\\
 1 & 4 & NA & NA & 2\\
 1 & 5 & 4.3168586 & 3 & 3\\
 1 & 6 & 5.1538039 & 1 & 1\\
 1 & 7 & NA& NA & 1 \\
 1 & 8 & 7.1767493 & 5 & 5\\
 1 & 9 & 8.9486051 & 1 & 1\\
 \hline
\end{tabular}
\caption{\small Example of continuous-time discretization of a single individual (ordered by time): the variables \textit{continuous time}  and  \textit{state} represent the simulated continuous-time points and their corresponding states. \textit{discrete time} denotes the fixed  discrete-time points from time $1$ to $T$, while \textit{LOCF state} represents the state values derived using the last observation carried forward (LOCF) method, computed from the continuous time. For instance, to generate values for all discrete-time points from the continuous-time data, note that no simulated observations exist at $t=3$ or $t=4$. Because the state transition following $t=1.98 $ does not occur until $t = 4.3$, the state value at $t=1.98$ (which is 2) is used to fill the state values for the discrete-time points 2, 3, and 4. The corresponding transition plot is illustrated in Figure ~\ref{fig:locf}}
\label{tab:generate_markov}
\end{table}

\subsection{Latent group estimation}
Three methods were applied to the simulation data: PCA with $K$-means clustering, LCA, and LTA. There are 200 simulation datasets, each of which contains 50 simulations. We then applied three methods to the simulated data. This simulation was done using \texttt{R}. 4.3.1. We begin by applying each approach to obtain the estimated latent class memberships $\hat{\bm{Z}}$. 
For the PCA-based method with $K$-means clustering, we introduce summary statistics inspired by SA at the end of Section ~\ref{sec:method}, and incorporate features that capture individual-level information over time, resulting in an input matrix of dimension $ N \times(p+1)$, where $p = 7$ denotes the number of summary statistics included, as described in Step 2 of Algorithm ~\ref{alg:PCA}. PCA was then applied to the proposed summaries above, and the first two PCs $(\bm{v_1}, \bm{v_2})$ were used as inputs for $K$-means clustering with $K=3$ to obtain the estimated latent class memberships $\hat{\mathbf{Z}}=\{\hat{Z}_{1},\dots, \hat{Z}_{N}\}$, where $\hat{Z}_i\in (1,\dots,K)$.
\vskip1ex
\begin{algorithm}[h]
\caption{PCA of Simulated Data}\label{alg:PCA}
\textbf{Given:} $K$ latent classes, and data $\bm{X}$ with $X_{it} \in \{1, \dots, C\}$, for $i=1,\dots,N$ individuals observed over $t=1, \dots T$ time points\;
\Indp
\textbf{1: Calculate summary statistics } \\
    \Indp
            \textbf{for} $i = 1$ to $N$\\
                \Indp
                Compute $\bar{X}_{i.} = \frac{1}{T} \sum_{t=1}^{T} X_{i,t}$\;
                Compute $SD(X_{i.})$\;
                Compute $\bar{L}_{i.} = \frac{1}{T-1} \sum_{t=2}^{T} L_{i,t}$\;
                Compute $SD(L_{i.})$\;
                Compute $P(L_{it}=0)= \frac{1}{T-1} \sum_{t=2}^{T-1} I(L_{i,t} = 0)$\;
                Compute $\text{Mode}_i = \arg\max_{c} \sum_{t=1}^{T} I(X_{i,t} = c)$\;
                Compute $\text{P(Mode}_i=X_{it}) = \frac{1}{T} \sum_{t=1}^{T} I(X_{i,t} = \text{Mode}_i)$\;
                \Indm  
            \vskip2ex
            \textbf{Output:} an $N \times (p+1)$ matrix, where $p = 7$, with rows for each individual:

     \begin{align*}
     \bm{X_{\text{sum}}}=
        \begin{bmatrix}
        1 & \bar{X}_{1.}  & SD(X_{1.}) & \bar{L}_{1.} & SD(L_{1.}) & P(L_{1t}=0) & Mode_1 & \text{P(Mode}_1=X_{1t}) \\
        \vdots & \vdots & \vdots & \vdots & \vdots & \vdots& \vdots&\\
        1 &\bar{X}_{N.}  & SD(X_{N.}) & \bar{L}_{N.} & SD(L_{N.}) & P(L_{Nt}=0)  & Mode_N & \text{P(Mode}_1=X_{1t}) 
        \end{bmatrix}\\
    \end{align*}
    \Indm\\[-.25in]
\textbf{2: Perform principal components analysis}\\
    \Indp
    use \texttt{stats::prcomp}, \textbf{input} $   \bm{X_{\text{sum}}}$
    \textbf{output:} principal components, $\bm{v_1}, \dots, \bm{v_{7}}$.\\
    \Indm
\textbf{3: Group principal components into latent classes using $K$-means}\\
    \Indp
    use \texttt{stats:kmeans} on $(\bm{v_1}, \bm{v_2})$ to generate estimated cluster membership, $\hat{\mathbf{Z}}=\{\hat{Z}_{1},\dots, \hat{Z}_{N}\}$, $\hat{Z}_i\in (1,\dots,K)$.\\
    \Indm
\Indm
\end{algorithm}

We then applied LCA, the function $\texttt{poLCA}$\cite{linzerPoLCAPolytomousVariable2026} was employed on the simulated data, with the number of classes set to $K = 3$ (see Algorithm ~\ref{alg:LCA}). The $\texttt{poLCA}$ package requires the input matrix $X$ to have dimensions $N \times (T+1)$ (wide format), where each element $X_{it}$ represents the observed value for individual $i$ at time $t$, with $i \in {1, \dots, N}$ and $t \in {1, \dots, T}$. The output consists of the estimated class memberships $\hat{\bm{Z}} = {\hat{Z}_1, \dots, \hat{Z}_N}$, where $\hat{Z}_i \in {1, \dots, K}$.

\begin{algorithm}[h]
\caption{LCA of Simulated Data}\label{alg:LCA}
\textbf{Given:} $K$ latent classes, and data $\bm{X}$ with $X_{it} \in \{1, \dots, C\}$, for $i=1,\dots,N$ individuals observed over $t=1, \dots T$ time points\;
\Indp
\textbf{1: Estimate latent classes via LCA } \\
    \Indp
            \textbf{Input:} an $N \times (T+1)$ wide format matrix of $\bm{X}$, with columns for each time point:\\
            \begin{center}
            \begin{tabular}{ c c c c }
             $i$ & $1 $ & $\dots$ & $T$ \\
             \hline
             1 & $X_{11}$ & $\dots$ & $X_{1T}$\\
             $\vdots$ & $\vdots$ & $\vdots$ & $\vdots$ \\
             N & $X_{N1}$ & $\dots$ & $X_{NT}$
            \end{tabular}
            \end{center}
            \textbf{Calculate:} use \texttt{poLCA::poLCA()} to fit an LCA model with $K$ latent classes\\
            \textbf{Output:} estimated latent class memberships, $\hat{\mathbf{Z}}=\{\hat{Z}_{1},\dots, \hat{Z}_{N}\}$, $\hat{Z}_i\in (1,\dots,K)$.\\
    \Indm
\Indm
\end{algorithm}
\vskip2ex

Finally, we applied LTA to the simulated data, using the $\texttt{lmest()}$ function \cite{Latent_Markov_LMEST_R_LTA}. This results in an estimated state proportion matrix ($C\times K$), where each column represents the proportion of observations in each of the $C$ levels within a specific latent class, where
\begin{equation*}
  \bm{\hat{\phi_k}} = (\hat{\phi}_{1|k},\dots,\hat{\phi}_{C|k})
\end{equation*}
 represents the response probabilities for class $k$. Each element $\hat{\phi}_{c|k}=P(X=c|S=k)$ is the probability of observing response level $c$ given membership in latent class $k$, where $S$ denotes the latent class at the population level (i.e., overall class membership), in contrast to $Z{i}$, which represents the latent state for individual $i$ at time $t$.
We then define \( \bm{h}_i = (h_{i1}, \dots, h_{iC}) \), where \( h_{ic} = \frac{1}{T} \sum_{t=1}^T I(X_{it} = c) \) represents the empirical proportion of time that individual \( i \) reports at stress level \( c \). Next, we compute the $L_2$ norm between the two vectors, $ G_{ik} = ||\bm{\hat{\phi}_k} - \bm{h_i} ||_{L_2} $ for all three estimated latent classes, and assign each individual to the estimated class with the smallest $L_2$ norm, $ G_{ik^*} = \min_{\substack{\> k }} G_{ik}$, where $\hat{Z_i}=k^*$. The details can be found in Algorithm ~\ref{alg:LTA}.\\
\begin{algorithm}[h]
\caption{LTA  Simulated Data}\label{alg:LTA}
\textbf{Given:} $K$ latent classes, and data $\bm{X}$ with $X_{it} \in \{1, \dots, C\}$, for $i=1,\dots,N$ individuals observed over $t=1, \dots T$ time points\;

\Indp
\textbf{1: Fit LTA} \;
    \Indp
            \textbf{input:} Let $\bm{X}^L$ be an $NT \times 3$ long format matrix of $\bm{X}$, with a single column for time:\\
            \begin{center}
            \begin{tabular}{ c c c  }
             i & t & observation \\
             \hline
            1 & 1 &$X_{11}$\\
            \vdots & \vdots & \vdots\\
            1 & T & $X_{1T}$ \\
            \vdots & \vdots & \vdots\\
            N & T & $X_{NT}$
            \end{tabular}
            \end{center}
            
            \textbf{Calculate:}

            Use \texttt{LMest::lmest} to fit an LTA model with K latent classes.\\
            \textbf{output:} Estimated conditional response probabilities,  $\hat{\phi_k} = (\hat{\phi}_{1|k},\dots,\hat{\phi}_{C|k})_{1\times C}$ for $ k = 1, \dots, K$\\
    \Indm
\Indm

\Indp
\textbf{2: Group individuals into latent classes by $\hat{\phi}_k$}\;
    \Indp
        \textbf{Input:} a $1\times C$ vector, $\bm{\hat{\phi_k}} = (\hat{\phi}_{1|k},\dots,\hat{\phi}_{C|k})$.\\
        \Indp
            \textbf{for} $i$ in $1, \dots N$\\
            \Indp
                \textbf{for} $c$ in $1, \dots C$\\
                \Indp 
                    calculate $h_{ic} = \frac{\sum^T_{t=1}I(X_{it}=c)}{T}$ \\
                \Indm
                Calculate $ G_{ik} = ||\bm{\hat{\phi}_k} - \bm{h_i} ||_{L_2} $ \\
            Find \\
                $ G_{ik^*} = \min_{\substack{\> k }} G_{ik}$ \\
            Set $\hat{Z_i}=k^*$ \\
            \Indm
        \Indm
        \textbf{Output:} estimated latent class memberships, $\hat{\mathbf{Z}}=\{\hat{Z}_{1},\dots, \hat{Z}_{N}\}$, $\hat{Z}_i\in (1,\dots,K)$.\\

    \Indm
    
\Indm
        
\end{algorithm}

Based on the value of estimated latent class variable $\bm{\hat{Z}}$ from the three approaches above, we then estimate the Markov process to generate the estimate transition probability matrix for each latent class. Due to the random labeling in the simulated data, we need to relabel the estimated classes to their corresponding true classes before evaluating classification performance. The Frobenius norm (F-norm) \cite{F_norm_Book} is computed between the estimated transition matrix and the true transition matrices of each class to identify which estimated latent class is closest to the true class. The class with the minimum Frobenius norm value among all $3\times3$ values is first assigned to the corresponding true class. Meanwhile, the values related to the estimated latent class and the assigned true class (1 + 2 $\times$ 2) are removed, leaving four remaining values. The same procedure is subsequently applied to assign the second and third estimated latent classes to the true class. Moreover, model performance is evaluated using precision, recall, accuracy, and estimated to true class size ratios. The details can be found in Algorithm ~\ref{identical_process}. 


\begin{algorithm}[h]
\caption{Relabel and Evaluation}\label{identical_process}
\textbf{Given:} data $\bm{X}$ with true latent memberships, $Z$, and estimated $\hat{Z}$\\
\textbf{1: Calculate estimated transition matrices }\\

        \Indp
            Let $\bm{X}^{Z} = [ \> \bm{X}^L \> \bm{\hat{Z}} \> ]$ be the long format of $\bm{X}$ with $\hat{\bm{Z}}$ appended.\\
            \textbf{for}{ $k = 1,\dots,K$}\\
                \Indp 
                Define $\bm{X}^Z_k = \{ X^Z_{it} | \hat{Z}_i = k\}$, the set of observations for the $k$th estimated latent class\\
                Apply \texttt{cfda::estimate\_Markov} to $\bm{X}_k^Z$\\
                \Indm
            \textbf{Output:} a $C \times C$ matrix of estimated transition probabilities for the $k$th latent class, $\hat{L}_{k}$\\
        
            \Indm
\textbf{2: Relabel estimated clusters }\\
\Indp
    \textbf{A: Compare estimated and true latent classes}\\
    \Indp
        \textbf{for} $k = 1, \dots, K$, \textbf{for} $j = k+1, \dots K$ \\[.02in]
        \Indp
            Calculate $ F_{kj} = || \hat{L}_{k} - L_{j} ||_F $ \\

        \Indm
    \Indm 

    \textbf{B: Assign new labels}\\
    \Indp
        Initialize new labels $\tau_k = 0$ for $k=1, \dots, K$.\\
        \Indp
            \textbf{while} any $\tau_k=0$\\
            \Indp
                 Find 
                    $$ F_{l^*j^*} = \min_{\substack{l \in \{k|\tau_k=0\}, \\ \> j \in \{1:K\} \setminus \tau_k }} F_{lj}$$ \\
                Set $\tau_{l^*}=j^*$ \\
            \Indm
        \Indm
    \Indm
    \textbf{Output:} relabeled estimated latent classes, $\tau_k$ \\
\Indm

\textbf{3: Classification accuracy}\; 
    \Indp
        \textbf{for} $k=1, \dots, K$ \\
        \Indp
            $TP_k = \sum^N_{n = 1}I(\hat{Z}_i=k, Z_i=k) $\\
            $TN_k = \sum^N_{n = 1}I(\hat{Z}_i\ne k, Z_i \ne  k)$\\
            $FP_k = \sum^N_{n =1 }I(\hat{Z_i}=k, Z_i \ne k)$\\
            $FN_k = \sum^N_{n=1}I(\hat{Z_i}\ne k, Z_i=k)$\\
           \textbf{Precision$_k$} =$ \frac{TP_k}{TP_k+FP_k}$\\
           \textbf{Recall$_k$} = $\frac{TP_k}{TP_k+FN_k}$ \\
           
           \textbf{Accuracy$_k$} = $\frac{TP_k+TN_k}{TP_k+TN_k+FP_k+FN_k}$\\
            \textbf{Estimated-to-true class size ratios$_k$} = $\frac{\sum I(\hat{Z}_i = k)}{\sum I(Z_i = k)}=\frac{n_{\text{est}_k}}{n_{\text{true}_k}}$.\\
        \Indm
    \Indm
\Indm
\end{algorithm}
\subsection{Results}
\begin{figure}[h]
      \centering                        
      \includegraphics[width=1\textwidth]{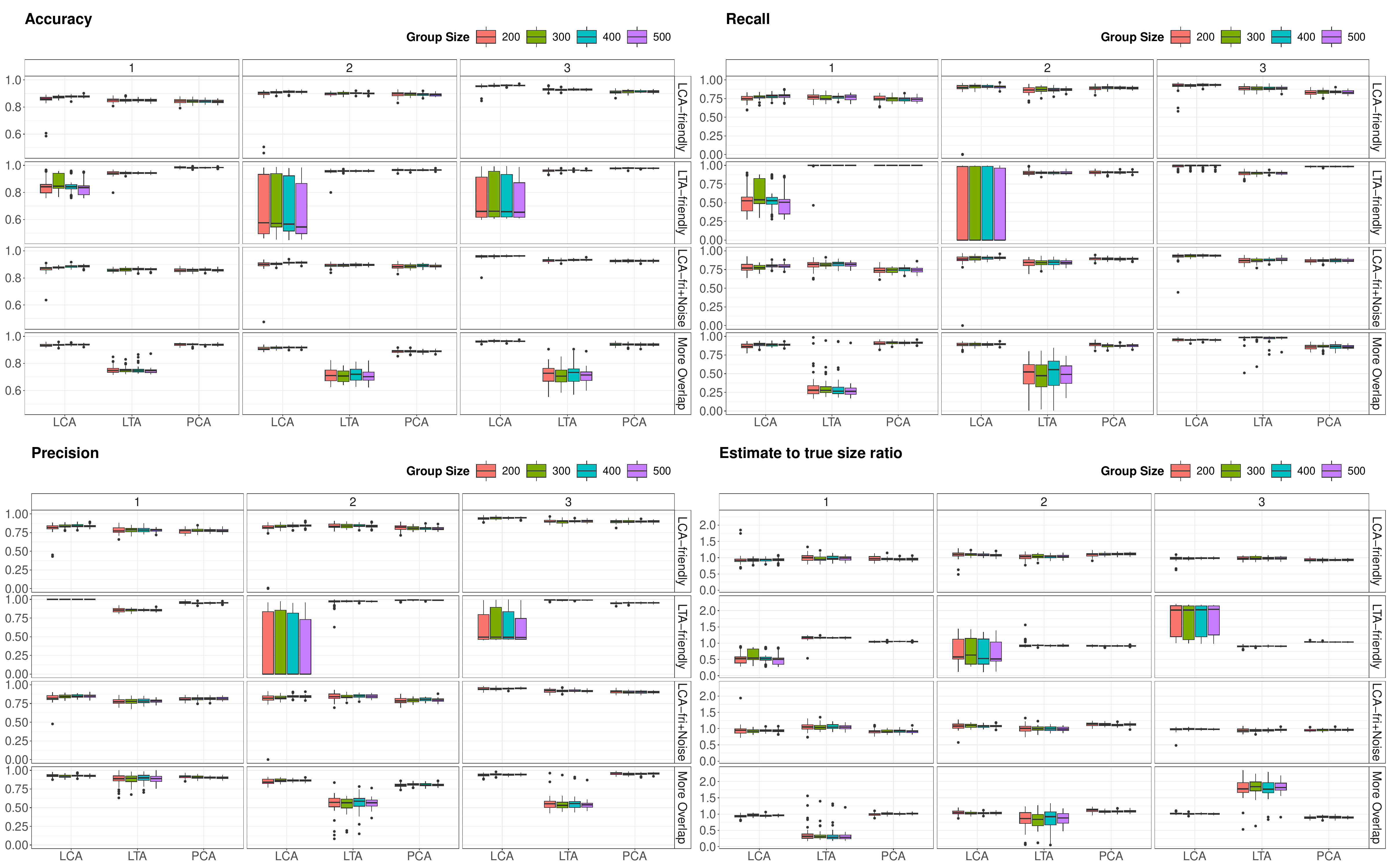}
      \caption[]{\small Simulation results: the multi-panel plot displays Recall, Accuracy, Precision, and the Estimate-to-True Size Ratio. Rows represent the four settings: LCA-friendly,  LTA-friendly, LCA-friendly+Noise, and More Overlap. Columns distinguish between the internal transition matrices ($L_1$, $L_2$, and $L_3$). Colors indicate group-specific sample size. }
      \label{fig: sim_res_all}
\end{figure}

We use a confusion matrix to interpret the model’s results, where the ideal values for precision, recall, and accuracy are all 1, indicating a potentially strong model for classification. Precision measures the proportion of correctly predicted positive cases among all predicted positives, and a low precision suggests too many false positives. Recall indicates the proportion of actual positives correctly identified by the model, and a low recall means many true positives are being missed. Accuracy reflects the overall proportion of correct predictions (both positive and negative), and a low accuracy implies generally poor model performance of overall classification \citet{Confusion_Matrix_2022}. The detailed results are presented in Figure ~\ref{fig: sim_res_all}.
\begin{table}[H]
\centering
\begin{tabular}{|p{4cm}|p{4cm}|p{4cm}|}
\hline
\textbf{} & \textbf{Actual Positive }  & \textbf{Actual Negative} \\ 
\textbf{} & \textbf{(Actual Class $k$)} & \textbf{(Actual not Class $k$)}\\
\hline
\textbf{Predicted Positive} & True Positive (TP) & False Positive (FP) \\ 
\textbf{ (Predicted Class $k$)} & \textbf{} & \textbf{} \\
\hline
\textbf{Predicted Negative} & False Negative (FN) & True Negative (TN) \\ 
\textbf{(Predicted not Class $k$)} & \textbf{} & \textbf{}\\
\hline
\end{tabular}
\caption{Confusion Matrix}
\label{tab:confusion_matrix}
\end{table}

In Setting 1, we keep the transition pattern very clear, with no large overlap between classes. All three methods produce very similar results across all group sizes. LCA shows generally better performance among all four criteria compared to LTA and PCA.  
In Setting 2, we created a loop pattern among all levels at class 1, as described above, which violates the local independence assumptions of LCA. It is clear that the recall of LCA in classes 1 and 2 shows a much wider range compared to other methods. In Group 1, the median bar is close to 0.5, while in Group 2, it is close to 0, with the upper quartile (Q3) approaching 1. Combining the results of accuracy, precision, and the estimate-to-true size ratio plot, we observe that LCA includes fewer individuals in Class 1 (only a portion of the true positives) and too many individuals in Class 2 (a large number of false positives). The results also show that PCA and LTA's perform very well.
In Setting 3, we increased the noise between the classes but did not change the overlap pattern. It is clear that all three methods performed very well under this condition.
In Setting 4, we increased the overlap between classes, which made it more difficult to distinguish the response variable levels across classes. From the results, we observe that LTA shows a wide range and low median recall in Classes 1 and 2 compared to LCA and PCA. The accuracy of LTA was also lower than that of LCA and PCA across all three classes.
In terms of precision, Class 1 had a similarly low median value and a wide boxplot compared to the other methods, while precision in Classes 2 and 3 was significantly lower than that of LCA and PCA.
Combining all the results, we see that LTA does not handle class overlap well, resulting in too many individuals being assigned to Class 3 and too few to Classes 1 and 2.
Overall, across all four settings, PCA demonstrates the most stable performance, with consistently comparable results, even in settings where all three methods perform well.

\section{EMA Measures of Stress and Memory Performance}\label{sec:realdata_app} 
We demonstrate our PCA-based approach using real data collected in the MyBPLab study \citep{MYBPLAB2023, MYBPLab2025}
to uncover potential common patterns of latent stress profiles across participants. 
Observations beginning after a gap of more than seven days were removed,  which corresponds to the  75th percentile of overall inter-observation gaps and ensures that at least 70\% of the cleaned data are included. As described in Section ~\ref{sec:data_cleaning}, the sample was further restricted to participants with at least 25 stress observations.
Unlike the previous simulation settings, establishing a universal upper limit for observations per person is unnecessary in this context. While LCA and LTA typically require temporal uniformity across subjects, our PCA-based approach accommodates individuals with measurements at varying time points. This flexibility is essential for the current dataset, which exhibits substantial variability in longitudinal density, with observations ranging from 25 to over 2,000 per individual.
The first two steps in Algorithm ~\ref{alg:PCA} were applied to a total of 3,194 subjects.  Informative summary statistics, inspired by sequence analysis were computed to capture complex temporal dynamics and subsequently summarized using PCA. Based on those representations, $K$-means clustering was then employed to estimate $K=3$ unobserved groups of individuals with similar stress profiles over time.

\subsection{Characterization of Stress Profiles Across Clusters}
\begin{figure}[h]
      \centering                        
      \includegraphics[width=0.6\textwidth]{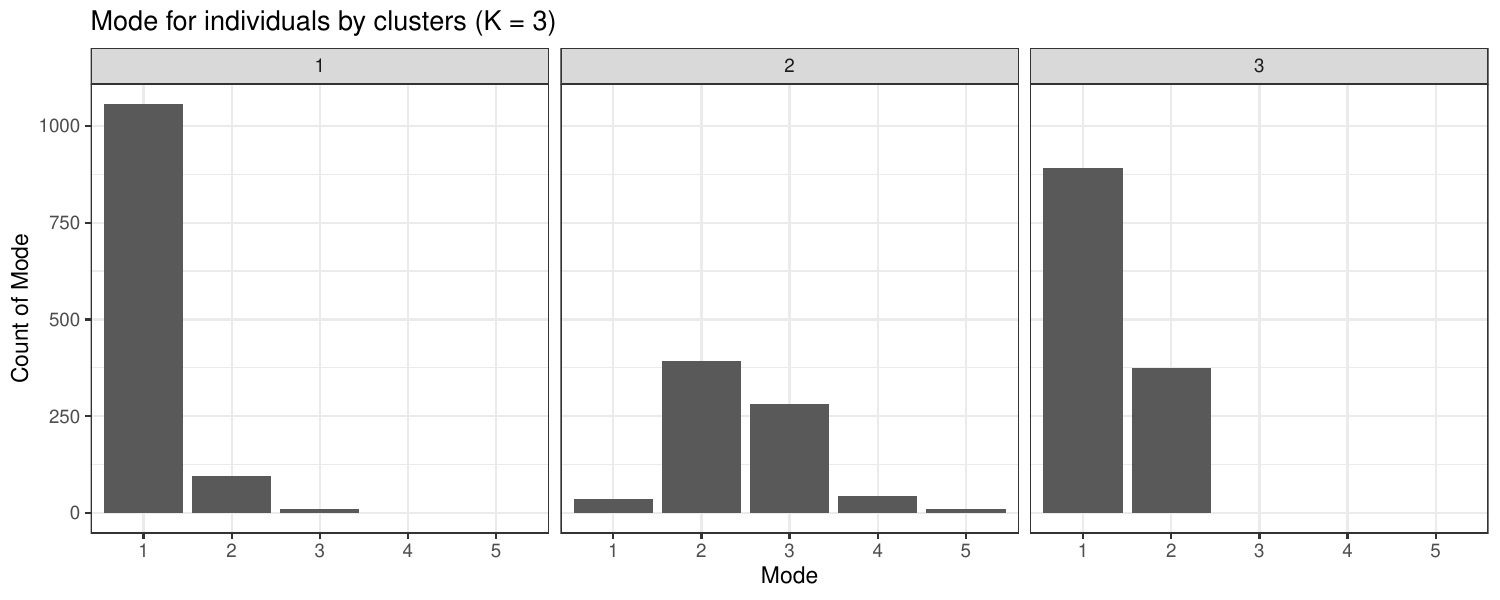}
    \caption{Distribution of individual mode by cluster: Each panel represents a latent cluster ($K=3$) estimated by our $PCA-$based approach, where the bars indicate the frequency of the most common response value (mode) for individuals within that group. }
      \label{fig:mode_k3}
\end{figure}

\begin{figure}[h]
      \centering                        
      \includegraphics[width=.6\textwidth]{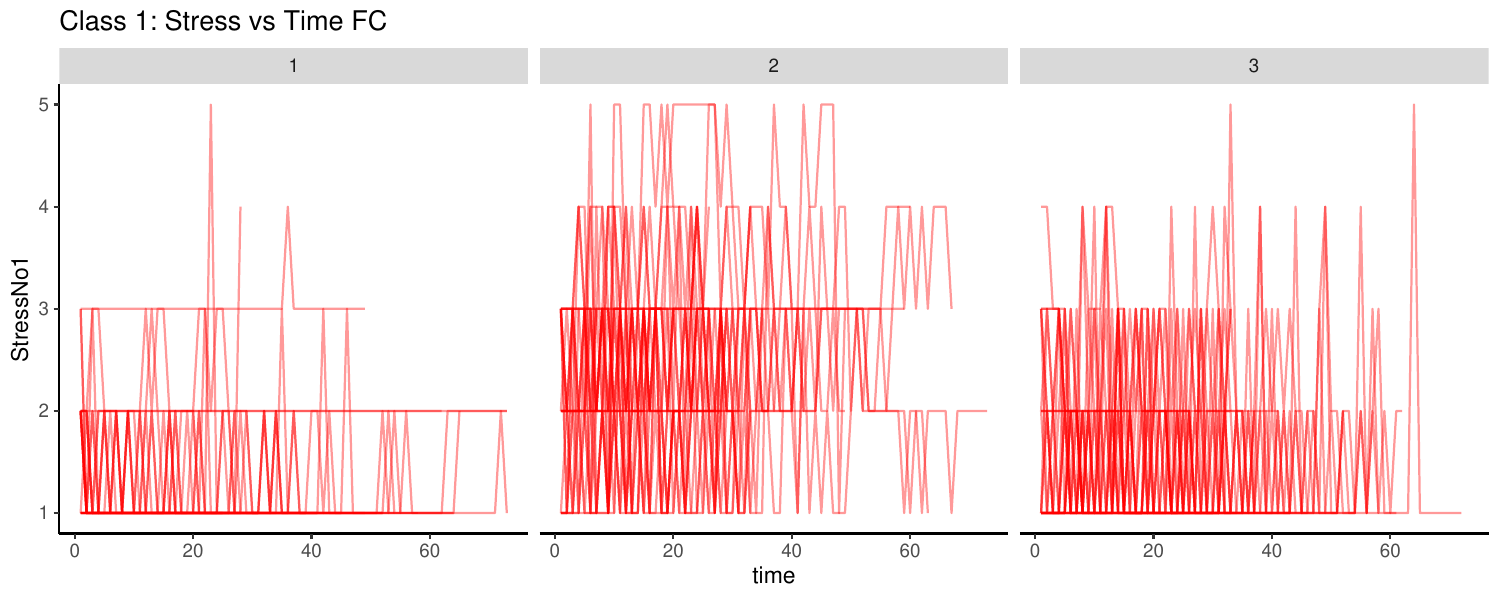}
        \includegraphics[width=.6\textwidth]{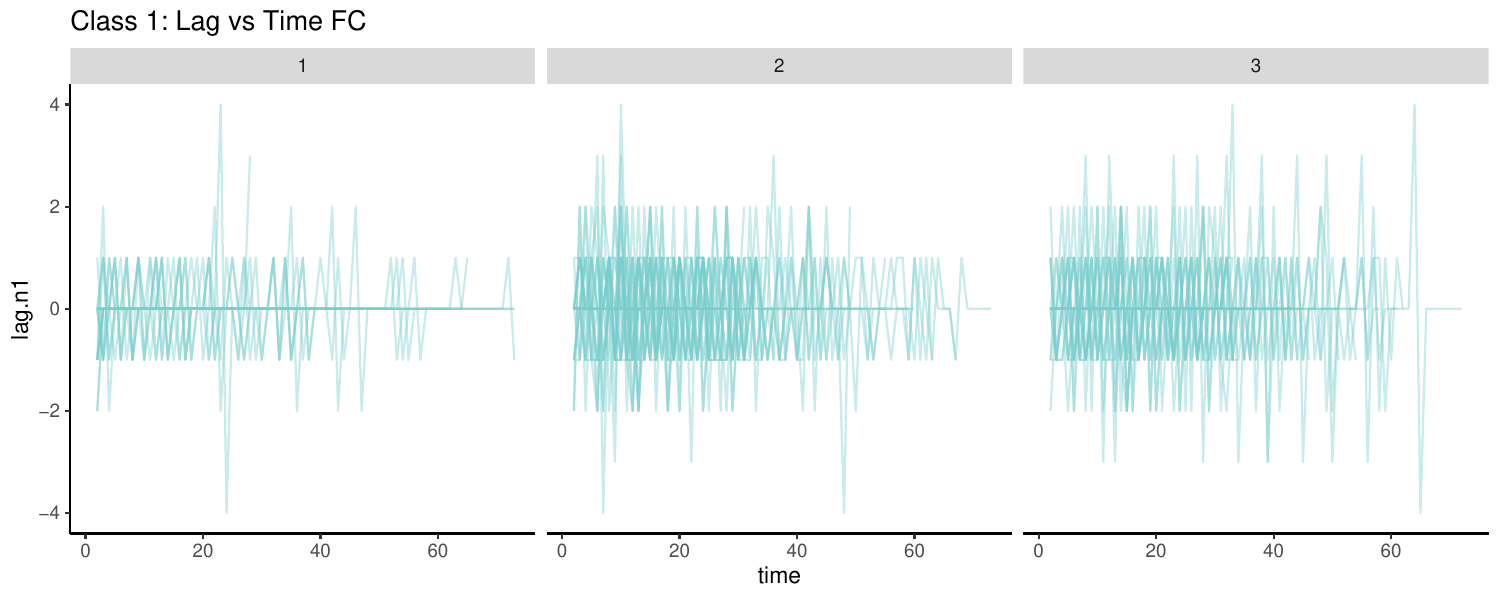}
     
      \caption[]{\small Longitudinal stress and lag trajectories by cluster.  Each column presents data for 25 randomly selected  individuals from each cluster ($K=3$), identified via our PCA-based approach. The top row illustrates reported stress levels over time, while the bottom row depicts change in consecutive stress score. To maintain visual interpretability and prevent scaling imbalances caused by high-frequency responders, individuals were randomly selected from those with fewer than 75 total stress records.}
       \label{fig:fuzzyClusterRES}
\end{figure}

To better understand the structure of these clusters, we examine their characteristics in terms of both informative summary statistics and demographic variables. Table ~\ref{tab:sum_k3_rus} presents the mean values of the informative summary statistics across the three clusters obtained from the $K$-means solution, along with their respective membership sizes.
Cluster 1 ($n=1,162$) represents the most stable group, characterized by the lowest mean stress (1.17) and the highest stability, with nearly 88\% of observations showing no change in stress level (mean of $P(L_{it}=0)$ = 0.88). 
In contrast, Cluster 2 ($n=765$) is the most volatile and high-stress group, exhibiting the highest mean stress (2.50) and the greatest variability in both stress scores (mean of $SD(X_{i.})=0.76$) and transition dynamics (mean of $SD(L_{i.})=0.91$). 
Cluster 3 ($n=1,267$) occupies an intermediate position with a mean stress of 1.61. While its probability of no change in stress level is higher than Cluster 2, it still shows more frequent fluctuations ($P(L_{it}= 0)$) than the highly stable Cluster 1. 
These patterns are further supported by Figure ~\ref{fig:mode_k3} and Figure ~\ref{fig:fuzzyClusterRES}, which illustrate additional cluster characteristics for the $K$-means approach when $K=3$. 

To further characterize these latent groups, we evaluated the demographic distributions presented in Table 3 in the supplementary document.
Cluster 1 is characterized by an older average age and a slightly lower proportion of White participants. In contrast, Clusters 2 and 3 include relatively younger individuals and a higher proportion of females compared to Cluster 1. Educational and behavioral differences are also present, as Cluster 1 has a higher proportion of individuals with a high school education or less, while Cluster 2 exhibits the highest levels of tobacco use. 
These variations suggest distinct demographic and behavioral profiles across the three groups, thereby confirming that the PCA-based clustering successfully partitions the population into groups with meaningful differences in longitudinal stress patterns.
\begin{table}[h]
\centering
\begin{tabular}{cccccccccc}
Clusters ($k$) & $SD(X_{i.})$ &$\bar{X}_{i.} $ &$\bar{L}_{i.}$ &$SD(L_{i.})$ &$P(L_{it}=0)$ &$\bar{T}_i$ &$\text{Mode}_i$ & $\text{P(Mode}_i=X_{it})$ &$n_k$\\
\hline
1 &0.27&	1.17&	0.00	&0.34&	0.88	&72.86&	1.10	&0.93	&1162\\
2 &0.76&	2.50	&0.02&	0.91&	0.49&	31.79&	2.48	&0.54	&765 \\
3 &0.68&	1.61&	-0.01&	0.87&	0.56&	36.22&	1.30&	0.64&1267 \\
\hline
\end{tabular}
\caption{Mean summary statistics by cluster. Each row displays the average values of the longitudinal features for a specific cluster ($K=3$). The columns correspond to the informative summary statistics described in Algorithm ~\ref{alg:PCA}, used to characterize the behavioral dynamics of each group.}
\label{tab:sum_k3_rus}
\end{table}

These patterns suggest a potential alignment with the CATS (low-stress) and GUTS (moderate-to-high stress) theoretical frameworks. Clusters 1 and 3, which exhibit lower overall stress levels (despite greater temporal variability in Cluster 3), align more closely with the CATS profile. In contrast, Cluster 2 reflects characteristics consistent with GUTS, including sustained moderate-to-high stress levels and more frequent state transitions. This interpretation remains exploratory, as the clustering was not explicitly guided by these theoretical frameworks.

\subsection{Mixed-Effects Models for Cognitive Performance Across Stress Clusters}
\begin{figure}[h] 
      \centering                        
      \includegraphics[width=1\textwidth]{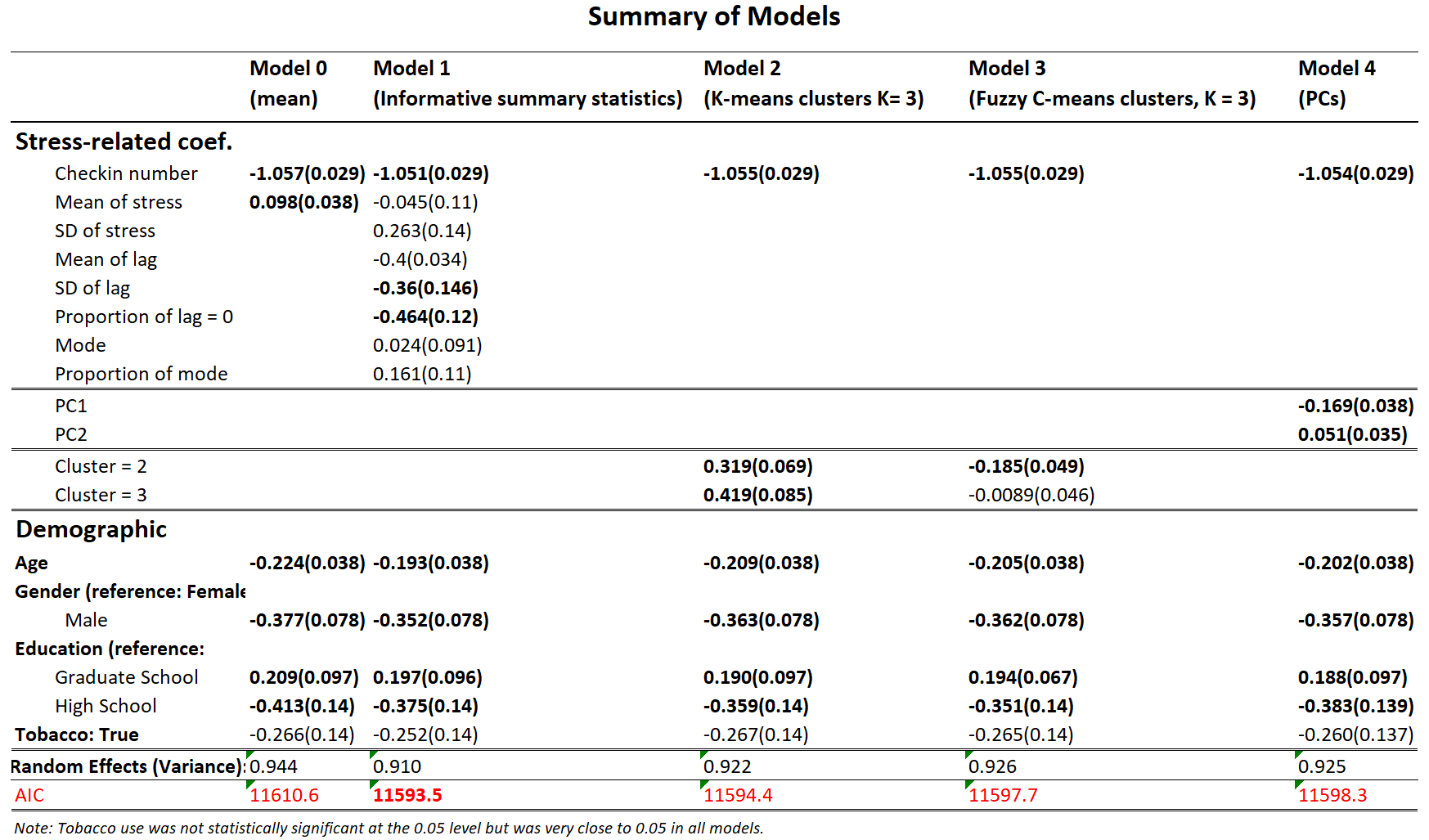}
      \caption[]{GLMM regression results across five model specifications. The table presents estimated coefficients and standard errors (in parentheses) for each model. Model 0 serves as the reference using mean stress. Model 1 incorporates the full set of informative summary statistics, while Model 2 represents the PCA-based K-means cluster. The sensitivity analysis are provided by Models 3 and 4 represents fuzzy ($C$-means) cluster memberships, and the first two PCs. respectively. All models include a standard set of demographic covariates and random effects, with the best-fitting model indicated by the minimum AIC value (highlighted in red).  }
      \label{tab:GLM_all}
\end{figure}

To examine the association between stress and memory outcomes and to evaluate the effects of the estimated latent classes, we estimated five versions of the generalized linear mixed effects model defined in Equation ~\ref{eq:glmer_model}. All models incorporated a standard set of demographic covariates described in Section ~\ref{sec: existing_app}. Race was coded as a series of binary indicators (Latino, White, Black, Asian, and Other), and the analysis was restricted to male and female participants due to sparse counts in other gender categories.
In each version, we replaced the individual's average reported stress, $\bar{X}_{i\cdot} $ with a model-specific variable, $ V_i $, which captures an individual's stress profiles. Model 0 serves as the reference model, utilizing only the individual’s mean reported stress as a predictor $ V_i = \bar{X}_{i\cdot} $.
We first focus on Model 2 which includes estimated latent cluster memberships from the PCA-based approach described earlier. $\pmb{V}_i$ is a categorical variable representing an individual’s membership in a specific cluster.
Coefficient estimates and model fit metrics are given in Table ~\ref{tab:GLM_all}, excluding non-significant demographic variables. Overall, Model 2, which includes the PCA-based latent clusters, achieves lower AIC values than the mean-only model (Model 0).  This indicates that our PCA-based approach provides a more informative summary of individuals' stress profiles over time, as they relate to cognitive performance.

We also checked the appropriate number of principal components by examining the scree plot. As shown in the scree plot (Figure 1 of the supplementary document
), the curve become relatively flat after the PC2, with the cumulative proportion of total variance explained reaching approximately 80\%. 


To further understand how clustering identifies different groups of stress profiles, we explore with different clustering approaches using both $K$-means clustering and Fuzzy $C$-means clustering. In $K$-means clustering, stress profile clusters cannot overlap and membership is strict (i.e., binary assignments).  In Fuzzy $C$-means clustering, stress profile clusters can overlap, meaning that individuals can belong to multiple clusters and share stress profile characteristics. Fuzzy $C$-means cluster assignments are not strict but are proportional to the likelihood of membership. To determine the optimal number of clusters, we calculated the silhouette score (SI)  \cite{SI_score_1987} for $K = $2, 3, and 4 using both $K$-means and Fuzzy $C$-means clustering (Table 1 in Supplementary document
). 
The highest mean SI was observed at $K=2$ for the $K$-means approach. In contrast, Fuzzy $C$-means showed only minimal differences in SI across different values of $K$. Since the results were not substantially different and were inconsistent between $K$-means and Fuzzy $C$-means, we proceed to fit models with $K = 2, \;3$, and $4$ for both clustering methods.
Comparison of AIC values across all configurations indicated that $K = 3$ yielded the lowest AIC among the clustering-based models, with the $K$-means setting providing the best overall fit. Model fit summaries using $K = 2$ and $4$ are provided in Table 2 from supplementary document.
Estimates from Model 3, which uses Fuzzy $C$-means cluster memberships are given in Table ~\ref{tab:GLM_all}. In this model, $\pmb{V}_i $ is a vector of length $K-1$ representing an individual's membership probabilities across the clusters, with the $K$th cluster excluded to ensure model identifiability.  

\begin{figure}[h] 
 \centering
      \hspace*{-1.5cm}             
      \includegraphics[width=0.7\textwidth]{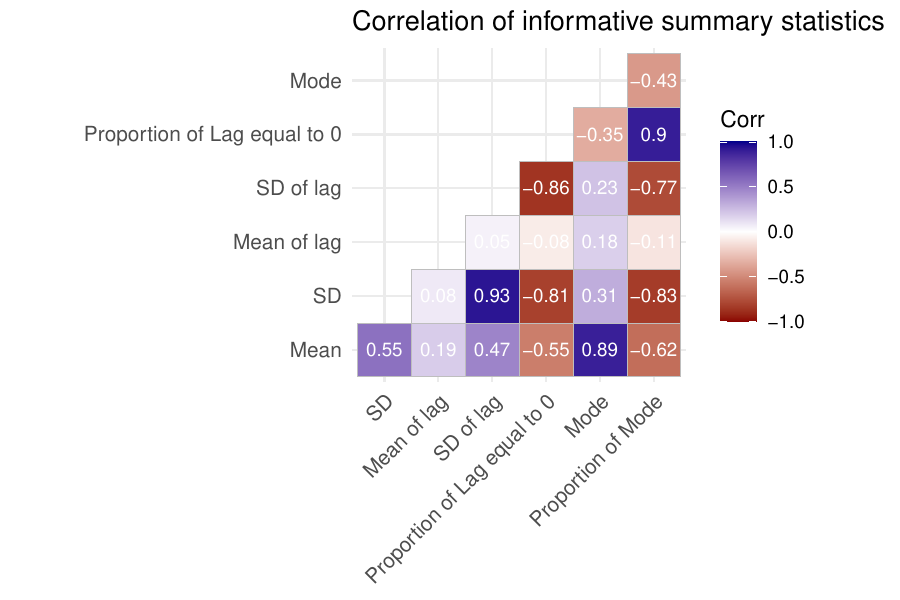}
      \caption[]{Correlation of informative summary statistics: This heatmap displays the pairwise Pearson correlation coefficients among the informative summary statistics derived from longitudinal stress trajectories. Color intensity reflects the strength and direction of the correlation, with darker shades indicating stronger positive (blue) or negative (red) associations. The $x$- and $y$-axes represent the different summary statistics included in the analysis.}
      \label{fig:infor_corr}
\end{figure}

As further sensitivity analysis, we considered alternative model specifications. In Table ~\ref{tab:GLM_all}, Models 1 and 4 use the summary statistics derived from Algorithm ~\ref{alg:PCA} and their associated principal components, respectively, as predictors to represent individual-level stress trajectories. Among all models, the specification incorporating all informative summary statistics (Model 1) achieved the lowest AIC (11593.5). 
However, it is generally recommended that variables with an absolute correlation greater than 0.7 not be used jointly, a guideline that is consistent with the simulation results reported in \citet{2013_collinerarity}. High collinearity indicates potential risks of redundant or shared information among predictors. Although model coefficients remain estimable, collinearity inflates the variance of parameter estimates and can lead to unstable and unreliable inference \cite{Wheeler_2007_collinearity}. Figure ~\ref{fig:infor_corr} indicates substantial multicollinearity among several summary statistics as further supported by the component loadings shown in Figure 2 in supplementary document
, where multiple pairs of variables contribute in similar directions and with comparable strengths. Due to this issue, Model 1 is less desirable. Under this consideration, we emphasize our proposed PCA-based $K$-means clustering approach as a more robust alternative.

Overall, the PCA-based K-means model provided the best results (AIC = 11,594.4) and avoids issues with multicollinearity. The estimated clusters revealed distinct stress profiles. Cluster 1 tended to remain at lower stress levels with less frequent switching. Cluster 2 exhibited outcomes distributed across all five stress levels with most observations concentrated in 2 and 3 and showed a 50\% probability of switching status over time. Cluster 3 was similar to class two but with most outcomes remaining in stress level 1 and 2.

\section{Discussion and conclusion}\label{sec:disc}
In this paper, we proposed a PCA-based approach that borrows tools from sequence analysis to uncover shared latent profiles among longitudinal ordinal data, as is common in the EMA setting.  Using simulated data, we compared our approach with two commonly used models for identifying underlying unobserved group structures. We found that while LTA and LCA performed well under certain conditions, the PCA approach combined with $K$-means clustering produced the most stable and consistent results across all scenarios. Moreover, the PCA-based approach accommodates data with varying numbers of repeated measurements, which is common in EMA data.

In our application, we estimated three distinct stress profiles, with Clusters 1 and 3 reflecting relatively lower overall stress levels, although Cluster 3 shows greater within individual variability over time, while Cluster 2 captures individuals with more sustained moderate to high stress and more frequent transitions between states. These patterns highlight meaningful differences in both stress levels and temporal dynamics across individuals.
Using these estimated shared stress profiles, we describe the relationship between stress and cognitive performance, in a standard regression framework.  We observe improved model fit over a model that includes only aggregate mean stress.  Our approach incorporates longitudinal trends by borrowing measures from sequence analysis and avoids issues of multicollinearity.

While incorporating stress profile cluster memberships improves model fit, the improvement is small, likely due to the noise and complexity inherent in real-world EMA data. This suggests that the current dataset may not contain clearly distinguishable unobserved groups related to stress, or that the relationship between stress profiles and memory performance is not particularly strong based on the available variables. On the other hand, the memory assessments were not consistently completed according to the scheduled testing protocol, resulting in missing outcome data for some participants. 
As motivation for our study of stress profiles, we reviewed two competing theories of stress activation, CATS and GUTS.  While we used these theories to motivate different clusters of stress profiles within the study population, it is important to note that the GUTS model describes \textit{unconscious} experience, which may not translate to self-reports of perceived daily stress, contained in the data we analyzed here. 
Differentiating between these theories would require novel experiments that manipulate participants' perceived control and safety.

Our approach provides researchers with an accessible framework for leveraging the rich, fine-grained information produced by high-frequency longitudinal EMA data collection, enabling more nuanced insights into complex behavioral and contextual patterns.  
Several extensions could further improve the proposed modeling approach. A more robust data filtering process could be considered to determine an appropriate threshold for the minimum number of time points, as the current 7-day gap threshold may benefit from sensitivity analyses using alternative thresholds to evaluate the stability of results across different cutoff values. Additionally, rather than removing observations with large temporal gaps, an alternative strategy is to retain such data points while applying an observation-weighting scheme where weights are assigned based on the estimated proportion of noise contributed by each observation. The SA-derived summary statistics used to identify stress profile groups can also be adjusted based on the objective of the research. For instance, researchers who are more interested in the effect of large ordinal jumps in the response may consider supplementing the existing statistics with additional measures such as the proportion of large gap jumps or summary statistics computed separately before and after a jump event, in order to more precisely capture the dynamics of interest.

Beyond preprocessing, extensions to the model structure itself may also be warranted. One promising direction is to extend the existing Bayesian latent class analysis approach \citep{qiuTutorialBayesianLatent2022} by incorporating a linear link function of the time variable as a covariate, thereby enabling class membership estimation to account for time-related associations and improving the interpretation of longitudinal dynamics. Furthermore, addressing missing EMA data arising under different missingness mechanisms, such as completely at random or block missingness, would also be valuable, as each mechanism may require distinct modeling strategies to avoid biased inference. Additionally, cluster estimation could be improved by incorporating post-hoc cluster merging criteria, particularly in settings where the initial number of clusters is large or where one cluster contains a disproportionately small number of observations relative to the remaining groups, as merging such clusters may enhance interpretability without substantial loss of information. Finally, while this study adopts a discrete time variable to align with the settings of \citep{MYBPLAB2023, MYBPLab2025}, future work may consider incorporating continuous time as a covariate within the clustering framework, which could offer greater flexibility in capturing the temporal structure of longitudinal EMA data. Collectively, these extensions would strengthen both the preprocessing pipeline and the model structure, broadening the framework's applicability to EMA data and its capacity to characterize complex longitudinal dynamics.

\bibliography{main}

\begin{figure}[ht]
\centering
\includegraphics[width=0.5\textwidth]{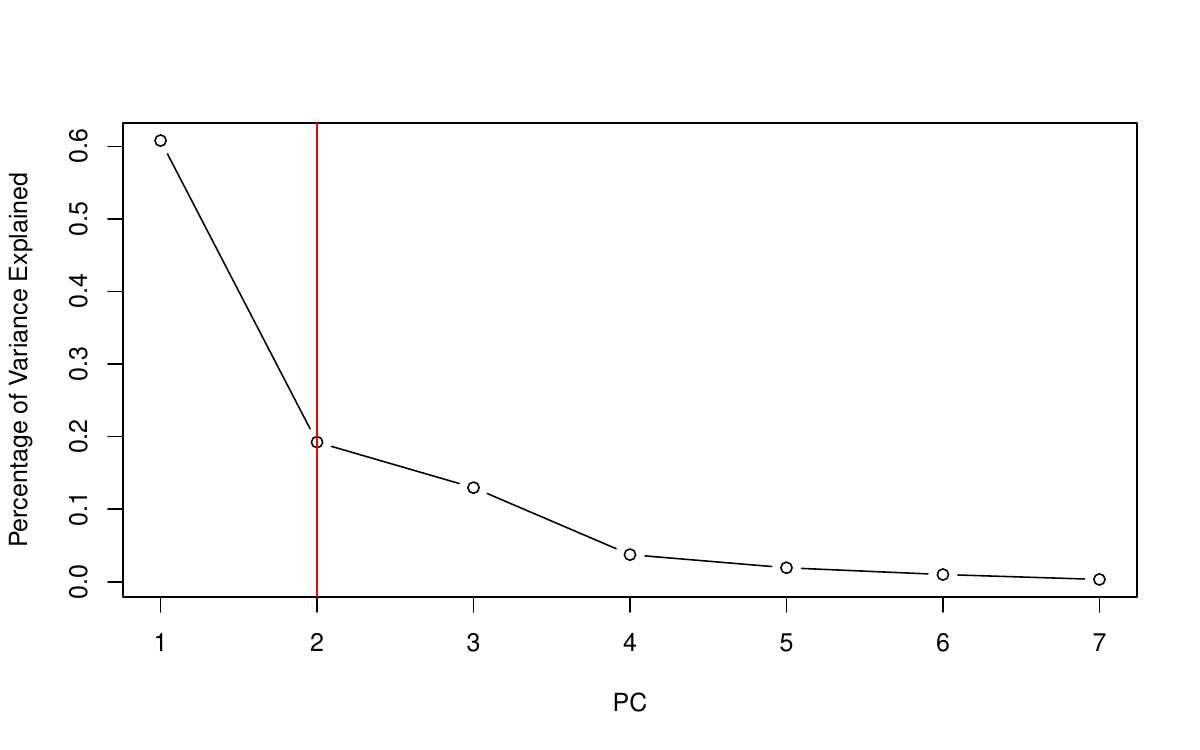}
\caption{\small Scree Plot of Principal Components. This plot displays the percentage of total variance explained by each principal component. A distinct "elbow" is observed at the second component, after which the marginal gain in explained variance reduces dramatically. }  
\label{fig:pca_scree} 
\end{figure}

\begin{table}[H]
\centering
\begin{tabular}{|c|c|c|c|}
\hline
    & $K = 2$ & $K = 3$ & $K = 4$ \\
\hline
    $K$-means & 0.3425 & 0.2568 & 0.2404 \\
\hline
   $C$-means & 0.2605 & 0.2263 &  0.2237 \\
\hline
\end{tabular}
\caption{\small Silhouette scores across varying values of $K$. Scores range from 0 to 1, with higher values indicating superior cluster separation and cohesion. The first row displays results for the $K$-means algorithm, while the second row corresponds to the $C$-means (fuzzy clustering) results.}
\label{tab:SI_score_table}
\end{table}

\begin{figure}[h] 
\centering
\includegraphics[width=0.6\textwidth]{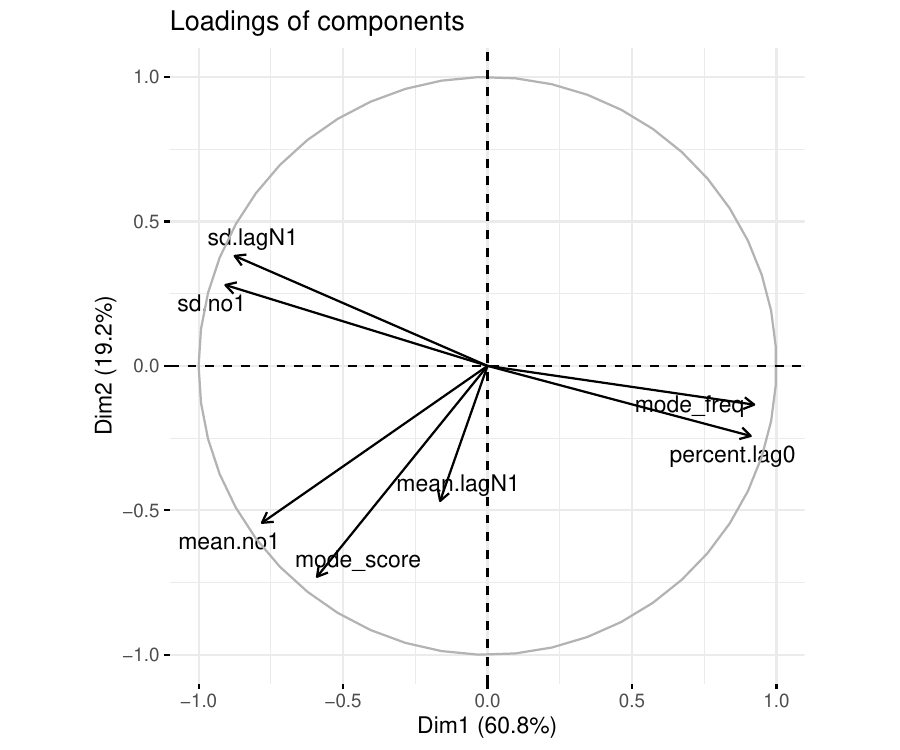}
\caption{\small Principal Component Loading Plot. This factor map illustrates the relationship between the informative summary statistics defined in Algorithm 1 
of the main text.
and the first two principal components. PC1 (Dim1, 60.8\%) and PC2 (Dim2, 19.2\%) together capture 80\% of the total variance. The vectors represent the direction and strength of each statistic's contribution to these dimensions.}  
\label{fig:PCAEllipses}
\end{figure}
\begin{table}[h]
      \centering                        
      \includegraphics[width=0.9\textwidth]{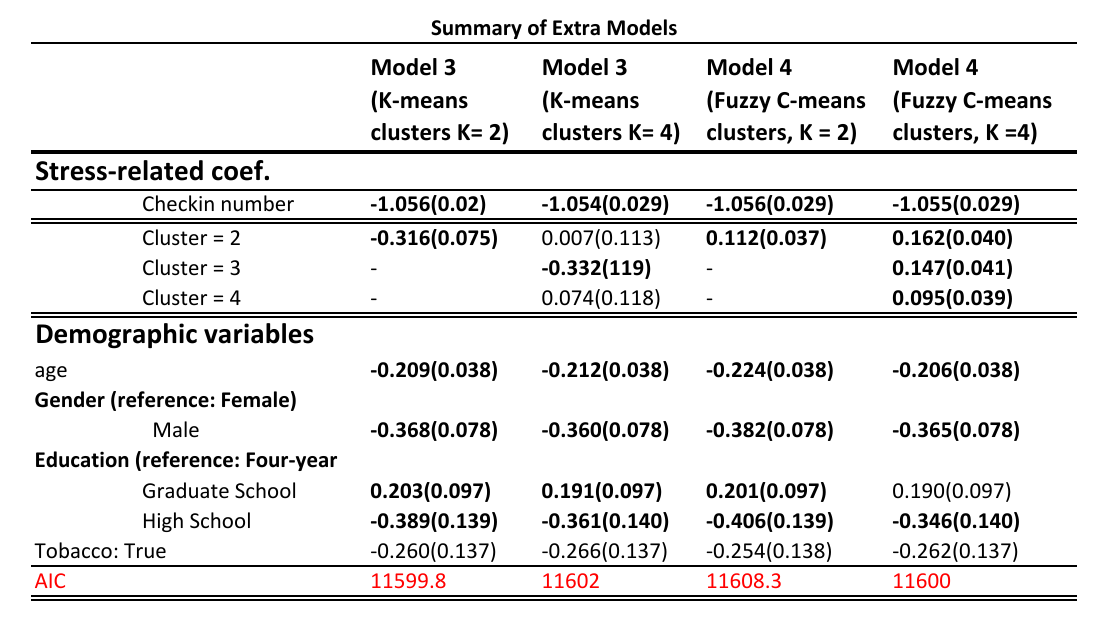}
      
      \caption[]{Extra GLMM Results \footnotemark }
      \label{GLM_extra_k_2_4}
\end{table}

\begin{table}[h]
      \centering                        
      \includegraphics[width=0.9\textwidth]{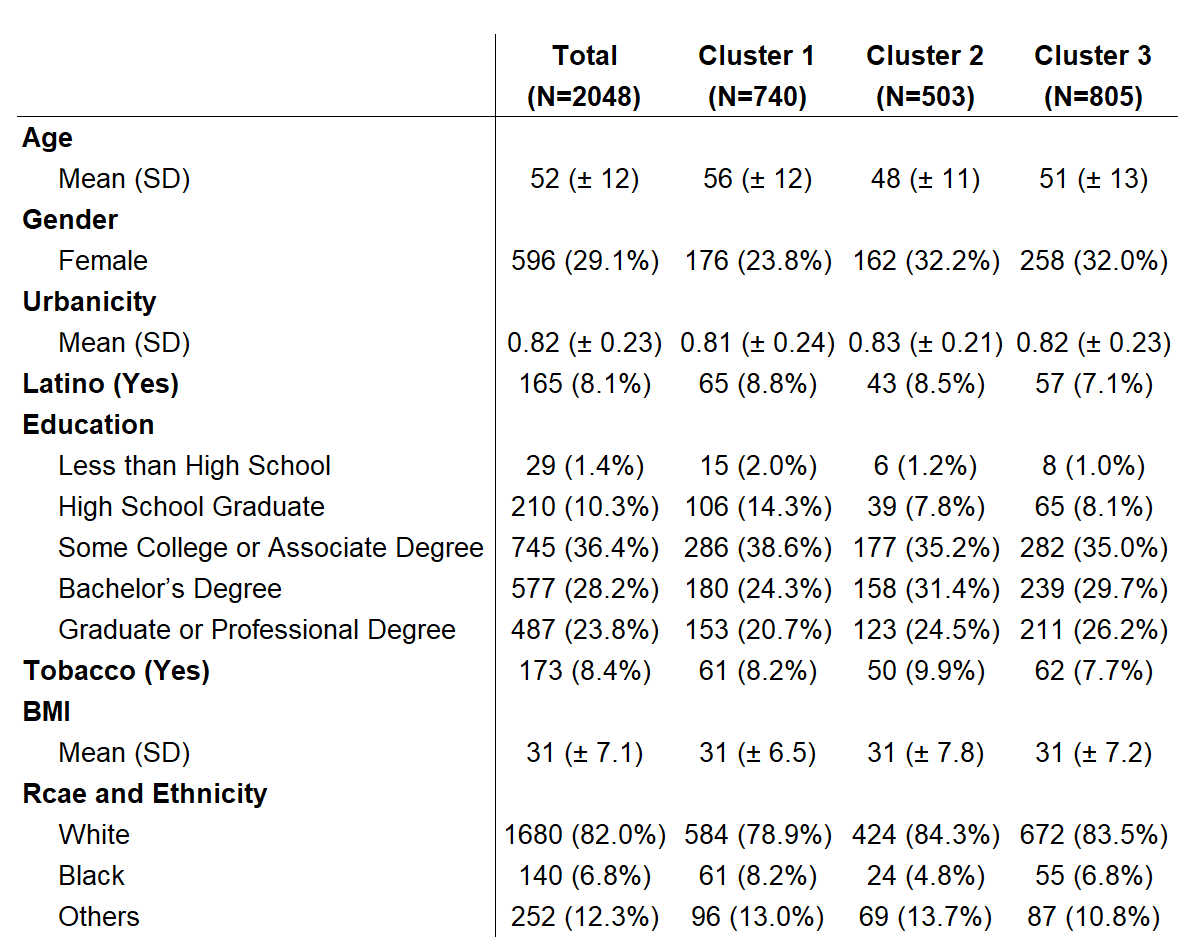}
      
      \caption[]{Demographic table: Summaries reported above describe the subset of individuals with non-missing values for the selected variables. Missingness for most demographic variables ranges from 16.5\% to 18.2\%, while urbanicity has a higher missing rate of approximately 35\%. Results are grouped by clusters identified using K-means (K = 3). Participants may report multiple race categories.}
      \label{demo_table}
\end{table}

\end{document}